\documentclass[a4paper,10pt]{article}
\usepackage[utf8]{inputenc}
\usepackage{hyperref}
\usepackage{amsfonts}
\usepackage{amsmath}
\usepackage{amssymb}
\usepackage{indentfirst}
\usepackage{graphicx}
\usepackage{cite}
\usepackage{color}
\usepackage{xcolor}

\usepackage{breqn}

\usepackage{setspace}%

\hypersetup{
colorlinks=true,
citecolor=black,
linkcolor=black,
filecolor=black,
urlcolor=black,
}
\numberwithin{equation}{section}
\allowdisplaybreaks
\makeatletter

\begin{document}

\begin{titlepage}
\rightline{MPP-2021-123}

\vskip 4cm

\begin{center}
\textbf{\LARGE{Gauge-invariant theories and higher-degree forms}}
\par\end{center}{\LARGE \par}

\begin{center}
	\vspace{1cm}
	\textbf{S. Salgado}
	\small
	\\[6mm]
	\textit{Max-Planck-Institut f\"{u}r Physik, F\"{o}hringer Ring 6, 80805 Munich, Germany }\\[0.5mm]
	\textit{Ludwig-Maximilians-Universit\"{a}t M\"{u}nchen, Theresienstra\ss e 37, 80333 Munich, Germany}
\textit{ }
	 \\[3mm]
	\footnotesize
	\texttt{E-mail: salgado@mppmu.mpg.de}
	 \\[5mm]
	\today
 
	\par\end{center}
\vskip 20pt
\centerline{{\bf Abstract}}
\medskip
\noindent A free differential algebra is generalization of a Lie algebra in which the
mathematical structure is extended by including of new Maurer--Cartan equations
for higher-degree differential forms. In this article, we propose a
generalization of the Chern--Weil theorem for free differential algebras
containing only one $p$-form extension. This is achieved through a
generalization of the covariant derivative, leading to an extension of the
standard formula for Chern--Simons and transgression forms. We also study the
possible existence of anomalies originated on this kind of structure. Some properties and
particular cases are analyzed.

\end{titlepage}\newpage {} 

\noindent\rule{162mm}{0.4pt}
\tableofcontents
\noindent\rule{162mm}{0.4pt}

\section{Introduction}

Higher gauge theories are generalizations of the standard gauge theories
that involve higher-degree differential forms. In the simplest case, this
means introducing not only the usual one-form gauge connection but also a
two-form gauge connection and a three-form field-strength, describing the
parallel transport along surfaces. It is possible to continue such extension
to gauge fields of degree higher than two, describing parallel transport
along extended objects. In a higher gauge theory \cite{baez}, the gauge
potentials are locally represented as $p$-forms, whose corresponding $(p+1)$%
-form gauge curvatures allow the construction of action principles. The
corresponding field equations are able to describe the dynamics of extended
objects, such as $p$-branes, in a similar manner in which a standard gauge
theory describes the dynamics of point particles. An example of this is
found in $p$-form electrodynamics \cite{teit}, whose gauge symmetry is
described by the invariance under the transformation law $A\rightarrow
A^{\prime}=A+\mathrm{d}\varphi$ where the abelian gauge field $A$ is a $p$%
-form, and $\varphi$ is a $(p-1)$-form and the parameter of the
transformation. Another example of this type is the rank-$2$ abelian
Kalb--Ramond gauge field \cite{zwie}. A generalization to non-abelian
higher-degree gauge theories has been studied on Refs. \cite%
{savv1,savv2,savv3}. In particular, in Refs.~\cite{epjs,sav4,sav5,snpb},
gauge-invariants forms were found, similar to the usual Chern--Pontryagin
densities and their corresponding Chern--Simons (CS) forms for special cases.

From a physical point of view, it is interesting to note that a common
feature to loop quantum gravity and string theory is the generalization of
point particles to extended objects. It is then interesting to study the
possible role that higher gauge theory could take in both frameworks.

In 1980 R. D'Auria, P. Fr\'{e} and T. Regge \cite{cohom} found an algebraic
structure known as free differential algebra (FDA) or Cartan integrable
system that allows formulating supergravity in the superspace in a geometric
manner, representing the spacetime as a supermanifold. In 1982, R. D'Auria
and P. Fr\'{e} made use of such structure to unveil a hidden symmetry
algebra in eleven-dimensional supergravity, previously constructed by Cremmer, Julia and Scherk \cite{dauria1,cjs}. On the other
hand, from Ref. \cite{townsend} it is known that the D'Auria--Fr\'{e}
formulation of supergravity is a higher-order geometric formulation of the
Cartan supergeometry, where extended algebraic structures replace Poincar%
\'{e} Lie superalgebra.

First-order formulations of supergravity in six or more dimensions have a
field content that includes bosonic higher-degree differential forms. Such
field presence is a consequence of the consistence requirement of an equal
number of bosonic and fermionic degrees of freedom in supersymmetry \cite%
{castellaniB}. Since the field content of these theories cannot be encoded
in one-forms dual to the generators of a Lie group, a possible solution is
to replace the concept of group manifold used in the formulation of gravity
and supergravity theories for a manifold that inherently involves higher
degree forms. This led to introduce some mathematical structures in physics
such as FDAs. These generalize the Maurer--Cartan equations that describe a
Lie algebra but including higher-degree differential forms as potentials.
FDAs are the natural generalization of Lie algebras, and since they include
Lie algebras as subalgebras, they can be used to describe the field content
of higher-dimensional gravity and supergravity theories.

The aim of this work is to use the FDA considered on Refs. \cite{cast1,
cast2, cast3, cast4} to obtain a gauge-invariant density and its
corresponding CS form and study the presence of anomalies in this kind of
theory. This paper is organized as follows: In Section 2, we briefly review
free differential algebras and their gauging, focused on the particular case
which will be important in the results of this article. In Section~3, we
will propose a definition of covariant derivative that will be necessary for
the construction of invariant gauge theories. In Section 4 we introduce a
generalization of the gauge invariant density of Lie algebras that includes
the $p$-form of the already mentioned FDA and study the corresponding
invariant tensor conditions. Section 5 contains a generalization of the
Chern--Weil theorem with explicit expressions for transgression and CS
actions for non-abelian gauge theory whose gauge fields are a one-form and a
\thinspace$p$-form ($p\geq2$). In Section 6 we finish with a study on the
existence of gauge anomalies for higher gauge theory, for which is also
necessary to introduce new notation and study some mathematical properties
about invariant tensors for FDAs. There are also two appendices with some
useful properties and an application for gravity.

\section{Free differential algebras}

The dual formulation of Lie algebras provided by the Maurer--Cartan
equations can be naturally extended to $p$-forms. Let us consider a basis of
differential forms $\left\{ \Theta ^{A\left( p\right) }\right\} _{p=1}^{N}$
defined on a manifold $M$ with $N\geq 2$. Each quantity $\Theta ^{A\left(
p\right) }\in \Lambda _{p}\left( M\right) $ is a differential $p$-form and
the algebraic index $A\left( p\right) $ takes values on different sets
depending on the value of $p$. Since $\left\{ \Theta ^{A\left( p\right)
}\right\} _{p=1}^{N}$ is a basis, the exterior derivative $\mathrm{d}\Theta
^{A\left( p\right) }$ can be written in terms of the same basis. This allows
to write a set of Maurer--Cartan (MC) equations for a mathematical structure
called free differential algebra%
\begin{equation}
\mathrm{d}\Theta ^{A\left( p\right) }+\sum_{n=1}^{N}\frac{1}{n}%
C_{B_{1}\left( p_{1}\right) \cdots B_{n}\left( p_{n}\right) }^{A\left(
p\right) }\Theta ^{B_{1}\left( p_{1}\right) }\wedge \cdots \wedge \Theta
^{B_{n}\left( p_{n}\right) }=0.  \label{fda6}
\end{equation}%
The coefficients $C_{B_{1}\left( p_{1}\right) \cdots B_{n}\left(
p_{n}\right) }^{A\left( p\right) }$ are called generalized structure
constants and are the generalization of the structure constants of Lie
algebras to the case of FDAs. The nilpotent condition $\mathrm{d}^{2}\Theta
^{A\left( p\right) }=0$ leads to the corresponding generalized Jacobi
identity%
\begin{align}
\mathrm{d}^{2}\Theta ^{A\left( p\right) }& =-\sum_{n,m=1}^{N}\frac{1}{m}%
C_{B_{1}\left( p_{1}\right) \cdots B_{n}\left( p_{n}\right) }^{A\left(
p\right) }C_{D_{1}\left( q_{1}\right) \cdots D_{m}\left( q_{m}\right)
}^{B_{1}\left( p_{1}\right) }\Theta ^{D_{1}\left( q_{1}\right) }\wedge
\cdots \wedge \Theta ^{D_{m}\left( q_{m}\right) }\wedge \Theta ^{B_{2}\left(
p_{2}\right) }\wedge \cdots \wedge \Theta ^{B_{n}\left( p_{n}\right) } 
\notag \\
& =0.  \label{fda6b}
\end{align}%
We will restrict the analysis to a particular case that has been extensively
studied in Refs. \cite{cast1,cast2}, in which the FDA is given by a Lie
algebra with only one $p$-form extension, i.e., depending only on a $1$-form 
$\mu ^{A}$ and a $p$-form $B^{i}$%
\begin{align}
\Theta ^{A\left( 1\right) }& =\mu ^{A}, \\
\Theta ^{A\left( p\right) }& =p!B^{i}, \\
\Theta ^{A\left( \text{others}\right) }& =0,
\end{align}%
and therefore, reducing (\ref{fda6}) to a set of two MC equations\footnote{%
From now on we will omit the wedge product between differential forms.} \cite%
{cast1, cast2}%
\begin{align}
\mathrm{d}\mu ^{A}+\frac{1}{2}C_{BC}^{A}\mu ^{B}\mu ^{C}& =R^{A}=0,
\label{mu} \\
\mathrm{d}B^{i}+C_{Aj}^{i}\mu ^{A}B^{j}+\frac{1}{\left( p+1\right) !}%
C_{A_{1}\cdots A_{p+1}}^{i}\mu ^{A_{1}}\cdots \mu ^{A_{p+1}}& =R^{i}=0.
\label{B}
\end{align}%
Eq. (\ref{B}) shows how the FDA is an extension of the Lie algebra defined
by Eq. (\ref{mu}), achieved by the addition of a new MC equation for a $p$%
-form potential $B$ whose non-trivial structure is given by the presence of
a cocycle%
\begin{equation}
\Omega ^{i}=\frac{1}{\left( p+1\right) !}C_{A_{1}\cdots A_{p+1}}^{i}\mu
^{A_{1}}\cdots \mu ^{A_{p+1}}.
\end{equation}%
For consistency, this $\left( p+1\right) $-form must be covariantly closed.
Otherwise, the second exterior derivative on Eq. (\ref{B}) does not vanish
and the generalized Jacobi identity would not hold. Note that, if the
cocycle is covariantly exact, i.e., $\Omega =\nabla \varphi $, it would be
possible to write the Maurer--Cartan equation (\ref{B}) as $\nabla B^{i}=0$,
through a redefinition of the field $B^{i}\rightarrow B^{i}+\varphi $.
Therefore, in order to have a non-trivial structure for the extended
algebra, the cocycle must be covariantly closed but not covariantly exact.
This means that, given a Lie algebra, there are as many non-equivalent FDA
extensions as Chevalley--Eilenberg cohomology classes the Lie algebra has \cite{cohom,dauria1}.

To define gauge transformations using this algebra, we need to write the
complete set of diffeomorphism transformations on the FDA manifold. Such
diffeomorphisms are given by the Lie derivatives along all the possible
directions on the FDA manifold. We need then to define a regular Lie
derivative (as with Lie groups) and an extended one that determines the
transformation of the $p$-form using a $\left( p-1\right) $-form parameter $%
\varepsilon^{j}$%
\begin{equation}
\begin{tabular}{ll}
$\ell_{\varepsilon^{A}t_{A}}=\mathrm{d}i_{\varepsilon^{A}t%
_{A}}+i_{\varepsilon^{A}t_{A}}\mathrm{d},$ & $\text{(Lie derivative
along the }0\text{-form }\varepsilon^{A}t_{A}\text{),}$ \\ 
$\ell_{\varepsilon^{j}t_{j}}=\mathrm{d}i_{\varepsilon^{j}t%
_{j}}+i_{\varepsilon^{j}t_{j}}\mathrm{d},$ & $\text{(Lie derivative
along the }\left( p-1\right) \text{-form }\varepsilon ^{j}t_{j}%
\text{).}$%
\end{tabular}
\ \ \ \ \ \ \ \ \ \ 
\end{equation}
These derivatives are defined in terms of the contraction operators $%
i_{\varepsilon^{A}t_{A}}=\varepsilon^{A}i_{t_{A}}$, $%
i_{\varepsilon^{j}t_{j}}=\varepsilon^{j}i_{t_{j}}$ whose
action is defined in terms of the basis as follows%
\begin{align}
i_{t_{A}}\mu^{B} & =\delta_{A}^{B},\text{ \ \ \ \ }i_{t%
_{A}}B^{j}=0, \\
i_{t_{j}}\mu^{A} & =0,\text{ \ \ \ \ }i_{t%
_{j}}B^{i}=\delta_{j}^{i}.
\end{align}
Here, $t_{A}$ are the generators of the Lie algebra described by
the Eqs. (\ref{mu}) (subalgebra of the FDA). In the same way, $t%
_{j} $ correspond to a basis of vectors on the FDA manifold in the direction
of the $p$-forms, i.e., $t_{j}$ is dual to $B^{j}$ in the same way
in which $t_{A}$ is dual to $\mu^{A}$ \cite{cast3, cast4}. Applying the regular Lie
derivative $\ell_{\varepsilon^{A}t_{A}}$ on the gauge fields $%
\mu^{A}$ and $B^{i}$ we have%
\begin{align}
\ell_{\varepsilon^{A}t_{A}}\mu^{A} & =\mathrm{d}\varepsilon
^{A}+C_{BC}^{A}\mu^{B}\varepsilon^{C}+2R_{\text{ \ }BC}^{A}\varepsilon^{B}%
\mu^{C},  \label{diff1} \\
\ell_{\varepsilon^{A}t_{A}}B^{i} & =\left( R_{\text{ \ }%
Aj}^{i}-C_{Aj}^{i}\right) \varepsilon^{A}B^{j}+\left( \left( p+1\right) R_{%
\text{ \ }AA_{1}\cdots A_{p}}^{i}-\frac{1}{p!}C_{AA_{1}\cdots
A_{p}}^{i}\right) \varepsilon^{A}\mu^{A_{1}}\cdots\mu^{A_{p}}.  \label{diff2}
\end{align}
Applying the extended Lie derivative $\ell_{\varepsilon^{j}t_{j}}$
on the gauge fields $\mu^{A}$ and $B^{i}$ we have%
\begin{align}
\ell_{\varepsilon^{j}t_{j}}\mu^{A} & =\varepsilon^{j}R_{\text{ \ }%
j}^{A},  \label{diff3} \\
\ell_{\varepsilon^{j}t_{j}}B^{i} & =\mathrm{d}%
\varepsilon^{j}+C_{Aj}^{i}\mu^{A}\varepsilon^{j}-R_{\text{ \ }%
Aj}^{i}\mu^{A}\varepsilon ^{j}.  \label{diff4}
\end{align}
Eqs. (\ref{diff1}) - (\ref{diff4}) contain the complete set of
diffeomorphism transformations along all the independent directions of the
FDA manifold \cite{cast3, cast4}. Both transformation laws depend on the
parameters $\varepsilon^{A}$ and $\varepsilon^{j}$ of the transformations
and can be sumarized as follows%
\begin{equation}
\delta\mu^{A}=\mathrm{d}\varepsilon^{A}+C_{BC}^{A}\mu^{B}\varepsilon
^{C}+2R_{\text{ \ }BC}^{A}\varepsilon^{B}\mu^{C}+\varepsilon^{j}R_{\text{ \ }%
j}^{A},
\end{equation}
\begin{dmath}
\delta B^{i} =\mathrm{d}\varepsilon^{j}+C_{Aj}^{i}\mu^{A}\varepsilon
^{j}-R_{Aj}^{i}\mu^{A}\varepsilon^{j}+\left(  R_{Aj}^{i}-C_{Aj}^{i}\right)
\varepsilon^{A}B^{j}+\left(  \left(  p+1\right)  R_{AA_{1}\cdots A_{p}}%
^{i}\right.  \nonumber\\
\left.  -\frac{1}{p!}C_{AA_{1}\cdots A_{p}}^{i}\right)  \varepsilon^{A}%
\mu^{A_{1}}\cdots\mu^{A_{p}}.
\end{dmath}Note that these transformations also depend on the curvature of
the FDA manifold. It is possible to obtain a restricted version of such
transformations demanding horizontality of the curvatures in some directions
of the FDA manifold, i.e.,%
\begin{align}
i_{\varepsilon^{B}t_{B}}\left( R_{CD}^{A}\mu^{C}\mu^{D}\right) & =0,
\\
i_{\varepsilon^{i}t_{i}}\left( R_{j}^{A}B^{j}\right) & =0, \\
i_{\varepsilon^{B}t_{B}}\left( R_{Ak}^{j}\mu^{A}B^{k}\right) & =0,
\\
i_{\varepsilon^{i}t_{i}}\left( R_{Ak}^{j}\mu^{A}B^{k}\right) & =0,
\\
i_{\varepsilon^{B}t_{B}}\left( R_{A_{1}\cdots A_{p+1}}^{i}\mu
^{A_{1}}\cdots\mu^{A_{p+1}}\right) & =0.
\end{align}
The curvature forms admit a splitting $R^{A}=R_{1}^{A}+R_{2}^{A}$, $%
R^{i}=R_{1}^{i}+R_{2}^{i}$ where%
\begin{equation}
\begin{tabular}{ll}
$R_{1}^{A}=R_{BC}^{A}\mu^{B}\mu^{C},$ & $R_{2}^{A}=R_{j}^{A}B^{j},$ \\ 
$R_{1}^{i}=R_{A_{1}\cdots A_{p+1}}^{i}\mu^{A_{1}}\cdots\mu^{A_{p+1}},$ & $%
R_{2}^{i}=R_{Aj}^{i}\mu^{A}B^{j},$%
\end{tabular}%
\end{equation}
such that the horizontality conditions can be written as follows%
\begin{align}
i_{\varepsilon^{B}t_{B}}R_{1}^{A} & =i_{\varepsilon^{B}t%
_{B}}R_{1}^{i}=i_{\varepsilon^{B}t_{B}}R_{2}^{i}=0,  \label{iR1} \\
i_{\varepsilon^{j}t_{j}}R_{2}^{A} & =i_{\varepsilon^{j}t%
_{j}}R_{1}^{i}=0.  \label{iR2}
\end{align}
In a more convenient way, Eqs. (\ref{iR1}) and (\ref{iR2}) can be written as 
\begin{equation}
i_{\varepsilon}R^{A}=0,\text{ \ \ \ \ }i_{\varepsilon^{B}t%
_{B}}R^{i}=0.
\end{equation}
As happens with Lie groups, with these conditions, the diffeomorphisms
become gauge transformations%
\begin{align}
\delta\mu^{A} & =\mathrm{d}\varepsilon^{A}+C_{BC}^{A}\mu^{B}\varepsilon ^{C},
\label{d1} \\
\delta B^{i} & =\mathrm{d}\varepsilon^{i}+C_{Aj}^{i}\mu^{A}\varepsilon
^{j}-C_{Aj}^{i}\varepsilon^{A}B^{j}-\frac{1}{p!}C_{A_{1}\cdots
A_{p+1}}^{i}\varepsilon^{A_{1}}\mu^{A_{2}}\cdots\mu^{A_{p+1}}.  \label{d2}
\end{align}
Eq. (\ref{d1}) corresponds to the usual covariant derivative of the $0$-form
parameter. The second equation is the natural extension to the case of a $p$%
-form. Note that the transformation of $\mu^{A}$ is the same that appears in
the study of standard gauge theory and it depends only on $\varepsilon^{A}$.
On the other side, the transformation of $B^{i}$ depends on $\varepsilon^{A}$
and $\varepsilon^{i}$ \cite{cast3,cast4}.

\section{Covariant derivative}

In order to formulate a gauge theory involving $p$-forms whose invariance is
governed by an FDA, it is necessary to define a covariant derivative that
involves all the components of the connection. To find such derivative, it
is useful to consider the information provided by the transformation law for
the curvatures and the Bianchi identities.

Using Eqs. (\ref{d1}) and (\ref{d2}) it is possible to prove the following
relations%
\begin{align}
\delta R^{A} & =C_{BC}^{A}R^{B}\varepsilon^{C},  \label{delta1} \\
\delta R^{i} &
=C_{Aj}^{i}R^{A}\varepsilon^{j}-C_{Aj}^{i}\varepsilon^{A}R^{j}-\frac{1}{%
\left( p-1\right) !}C_{AA_{1}\cdots A_{p}}^{i}\varepsilon
^{A}R^{A_{1}}\mu^{A_{2}}\cdots\mu^{A_{p}}.  \label{delta2}
\end{align}
Eq. (\ref{delta1}) is equivalent to the Lie bracket between the $2$-form
curvature and the $0$-form parameter. The natural generalization for the $%
\left( p+1\right) $-form curvature is expressed in Eq. (\ref{delta2}). We
can see that $R^{i}$ also transforms homogeneously, i.e., not depending on
derivatives of the parameters.

On the other hand, starting from the definition of the curvature forms, we
can also calculate its exterior derivative to find the Bianchi identities%
\begin{align}
\mathrm{d}R^{A}-C_{BC}^{A}R^{B}\mu^{C} & =0,  \label{bianchi1} \\
\mathrm{d}R^{i}+C_{Aj}^{i}\mu^{A}R^{j}-C_{Aj}^{i}R^{A}B^{j}-\frac{1}{p!}%
C_{A_{1}\cdots A_{p+1}}^{i}R^{A_{1}}\mu^{A_{2}}\cdots\mu^{A_{p+1}} & =0.
\label{bianchi2}
\end{align}
Since Eq. (\ref{mu}) is the MC equation for a Lie algebra, it follows that
Eqs. (\ref{delta1}) and (\ref{bianchi1}) reproduce the variation of the
curvature and Bianchi identity for Lie groups. Such relations and their
corresponding extended versions are necessary to define the covariant
derivative. However, we can see that the gauge transformations and Bianchi
identities lead to different definitions of covariant derivatives.

From Eqs, (\ref{d1}) and (\ref{d2}) we can see that if we have a set of
differential forms $\varepsilon=\left( \varepsilon^{A},\varepsilon
^{i}\right) $, where $\varepsilon^{A}$ is a $0$-form and $\varepsilon^{i}$
is a $\left( p-1\right) $-form, then its covariant derivative is given by $%
\nabla\varepsilon=\left( \left( \nabla\varepsilon\right) ^{A},\left(
\nabla\varepsilon\right) ^{i}\right) $, with%
\begin{align}
\left( \nabla\varepsilon\right) ^{A} & =\mathrm{d}\varepsilon^{A}+C_{BC}^{A}%
\mu^{B}\varepsilon^{C},  \label{covariant1} \\
\left( \nabla\varepsilon\right) ^{i} & =\mathrm{d}\varepsilon^{i}+C_{Aj}^{i}%
\mu^{A}\varepsilon^{j}-C_{Aj}^{i}\varepsilon^{A}B^{j}-\frac{1}{p!}%
C_{AA_{1}\cdots A_{p}}^{i}\varepsilon^{A}\mu^{A_{1}}\cdots\mu^{A_{p}}.
\label{covariant2}
\end{align}
On the other hand, from Eqs. (\ref{bianchi1}) and (\ref{bianchi2}) we can
see that if we consider the curvature tensor of the FDA $R=\left(
R^{A},R^{i}\right) $ where $R^{A}$ is a $2$-form and $R^{i}$ is a $\left(
p+1\right) $-form, then its covariant derivative is given by $\nabla
R=\left( \left( \nabla R\right) ^{A},\left( \nabla R\right) ^{i}\right) $,
with%
\begin{align}
\left( \nabla R\right) ^{A} & =\mathrm{d}R^{A}+C_{BC}^{A}\mu^{B}R^{C},
\label{covariant3} \\
\left( \nabla R\right) ^{i} & =\mathrm{d}R^{i}+C_{Aj}^{i}%
\mu^{A}R^{j}-C_{Aj}^{i}R^{A}B^{j}-\frac{1}{p!}C_{A_{1}\cdots
A_{p+1}}^{i}R^{A_{1}}\mu^{A_{2}}\cdots\mu^{A_{p+1}}.  \label{covariant4}
\end{align}

A possible answer is that these equations define the covariant derivative of
every set of differential forms $x=\left( x^{A},x^{i}\right) $ in the $%
\left( A,i\right) $-representation. However, there are two problems with
this definition:

\begin{itemize}
\item It does not satisfy the homogeneity condition, i.e., the second
covariant derivative $\nabla^{2}x$ depends on the exterior derivative of $x$.

\item The gauge curvature does not satisfy $\delta R^{i}=\nabla\delta B^{i}$.
\end{itemize}

At this point, we only know how to take the derivative of $\varepsilon$ and $%
R$. To solve this caveat, it is necessary to find a general definition that
depends on the order of the differential forms of the corresponding set of
fields. In the first case we have a set $\left(
\varepsilon^{A},\varepsilon^{i}\right) $ of $\left( 0,p-1\right) $-forms and
in the second one we have set $\left( R^{A},R^{i}\right) $ of $\left(
2,p+1\right) $-forms. In general, we have arrays of $\left( q,p+q-1\right) $%
-forms, and therefore, a general covariant derivative must be defined in
terms of $q$. Let us introduce a $\left( q,p+q-1\right) $-set of
differential forms in the $\left( A,i\right) $-representation of the algebra 
$x=\left( x^{A},x^{i}\right) $. We propose that the covariant derivative of $%
x$ can be written in components as follows%
\begin{align}
\left( \nabla x\right) ^{A} & =\mathrm{d}x^{A}+C_{BC}^{A}\mu^{B}x^{C}, \\
\left( \nabla x\right) ^{i} & =\mathrm{d}x^{i}+C_{Aj}^{i}\left( \mu
^{A}x^{j}-\left( -1\right) ^{f\left( q\right) }x^{A}B^{j}\right) -\frac{%
\left( -1\right) ^{g\left( q\right) }}{p!}C_{A_{1}\cdots
A_{p+1}}^{i}x^{A_{1}}\mu^{A_{2}}\cdots\mu^{A_{p+1}},
\end{align}
where we have introduced the functions $f\left( q\right) $ and $g\left(
q\right) $ in the terms involving the gauge fields and structure constants.
From (\ref{covariant2}) and (\ref{covariant4}) we have%
\begin{equation}
\left( -1\right) ^{f\left( 0\right) }=\left( -1\right) ^{f\left( 2\right)
}=\left( -1\right) ^{g\left( 0\right) }=\left( -1\right) ^{g\left( 2\right)
}=1.
\end{equation}
Besides, in order to satisfy the homogeneity condition, we need to remove
the dependence on the derivatives of $x^{A}$ and $x^{i}$ in the second
covariant derivatives of $x$. Those requirements lead to the following
conditions%
\begin{equation}
\left( -1\right) ^{f\left( q\right) }+\left( -1\right) ^{f\left( q+1\right)
}=0,\text{ \ \ \ }\left( -1\right) ^{g\left( q\right) }+\left( -1\right)
^{g\left( q+1\right) }=0,
\end{equation}
from which we have%
\begin{align}
f\left( q+1\right) & =f\left( q\right) +1, \\
g\left( q+1\right) & =g\left( q\right) +1.
\end{align}
A valid solution to this equation system is given by $f\left( q\right)
=g\left( q\right) =q$, giving us the following definition%
\begin{align}
\left( \nabla x\right) ^{A} & =\mathrm{d}x^{A}+C_{BC}^{A}\mu^{B}x^{C},
\label{def1} \\
\left( \nabla x\right) ^{i} & =\mathrm{d}x^{i}+C_{Aj}^{i}\left( \mu
^{A}x^{j}-\left( -1\right) ^{q}x^{A}B^{j}\right) -\frac{\left( -1\right) ^{q}%
}{p!}C_{A_{1}\cdots A_{p+1}}^{i}x^{A_{1}}\mu^{A_{2}}\cdots\mu^{A_{p+1}}.
\label{def2}
\end{align}
Note that this definition satisfies the homogeneity condition, i.e., the
second covariant derivative of $x$ does not depend on the derivatives on $x$%
. With this definition, we can write the variation of the gauge curvatures as%
\begin{align}
\delta R^{A} & =\nabla\delta\mu^{A}, \\
\delta R^{i} & =\nabla\delta B^{i}.
\end{align}

\section{Invariant density}

Now we postulate an invariant density, analogous to the Chern--Pontryagin
invariant of a Lie group. Using combinations of the curvavure forms, the
most general $q$-form that can be written in terms of $R^{A}$ and $R^{i}$ is
given by\footnote{%
We define $q\left( p\right) $ as the set of non-negative integer solutions $%
\left( m,n\right) $ to the algebraic equation $2m+\left( p+1\right) n=q$.}%
\begin{equation}
\chi_{q}=\sum_{m,n\in q\left( p\right) }g_{A_{1}\cdots A_{m}i_{1}\cdots
i_{n}}\underset{2m\text{-form}}{\underbrace{R^{A_{1}}\cdots R^{A_{m}}}}%
\underset{\left( p+1\right) n\text{-form}}{\underbrace{R^{i_{1}}\cdots
R^{i_{n}}}}.  \label{inv}
\end{equation}
The constants $g_{A_{1}\cdots A_{m}i_{1}\cdots i_{n}}$ must be such that $%
\chi_{q}$ is gauge invariant, namely $\delta\chi_{q}=0$. The sum runs over
all the possible values of $m$ and $n$ such that $2m+\left( p+1\right) n=q$,
and hence the resulting form is a $q$-form. Note the coefficients $%
g_{A_{1}\cdots A_{m}i_{1}\cdots i_{n}}$ contain mixed indices $A$ and $i$.
Enforcing the gauge invariance condition $\delta\chi_{q}=0$ will constraint
the form of such coefficients. Notice that the total variation of $\chi_{q}$
under gauge transformations is given by 
\begin{dmath}
\delta\chi_{q} =\sum_{2m+\left(  p+1\right)  n=q}g_{A_{1}\cdots A_{m}%
i_{1}\cdots i_{n}}\left[  mC_{BC}^{A_{1}}R^{B}\varepsilon^{C}R^{A_{2}}\cdots
R^{A_{m}}R^{i_{1}}\cdots R^{i_{n}}\right.  +nR^{A_{1}}\cdots R^{A_{m}}\left(
C_{Bj}^{i_{1}}R^{B}\varepsilon^{j}\right. \nonumber\\
\left.  \left.  -C_{Bj}^{i_{1}}\varepsilon^{B}R^{j}-\frac{1}{\left(
p-1\right)  !}C_{BB_{1}\cdots B_{p}}^{i_{1}}\varepsilon^{B}R^{B_{1}}\mu
^{B_{2}}\cdots\mu^{B_{p}}\right)  R^{i_{2}}\cdots R^{i_{n}}\right]  .
\end{dmath}Since the parameters $\varepsilon^{A}$, $\varepsilon^{j}$ are
independent, the terms proportional to each one have to vanish independently
without imposing any extra condition on the fields and curvatures. This
allows to split the independent conditions, resulting in the following three
equations%
\begin{align}
\sum_{r=1}^{m}C_{A_{0}A_{r}}^{C}g_{A_{1}\cdots\hat{A}_{r}C\cdots
A_{m}i_{1}\cdots i_{n}}+\sum_{s=1}^{n}C_{A_{0}i_{s}}^{k}g_{A_{1}\cdots
A_{m}i_{1}\cdots\hat{\imath}_{s}k\cdots i_{n}} & =0,  \label{it1} \\
\sum_{r=1}^{m+1}C_{A_{r}B_{1}\cdots B_{p}}^{i_{1}}g_{A_{1}\cdots\hat{A}%
_{r}\cdots A_{m+1}i_{1}\cdots i_{n}} & =0,  \label{it2} \\
\sum_{r=1}^{m+1}C_{A_{r}j}^{i_{1}}g_{A_{1}\cdots\hat{A}_{r}\cdots
A_{m+1}i_{1}\cdots i_{n}} & =0,  \label{it3}
\end{align}
where the indices with hat $\hat{A}$ and $\hat{\imath}$ denote the absence
of $A$ and $i$ in the sequence. Eqs. (\ref{it1}) - (\ref{it3}) correspond to
the definition of extended invariant tensor for the FDA. Note that for $n=0$
(this is, in the absence of $p$-form extension), Eq. (\ref{it1}) is
equivalent to the standard definition of the invariant tensor of a Lie
algebra. A short calculation shows that if the quantities $g_{A_{1}\cdots
A_{m}i_{1}\cdots i_{n}}$ satisfy the invariant tensor conditions (\ref{it1})
- (\ref{it3}), then $\chi_{q}$ becomes a closed form, i.e., $\mathrm{d}%
\chi_{q}=0$.

\subsection{Adjoint representation}

An interesting case can be found when the $p$-form $B^{i}$ is also in the
adjoint representation of the Lie subalgebra. In such case, the structure
constants $C_{Ak}^{i}$ become equivalent to the structure constants of the
Lie subalgebra, i.e.,%
\begin{equation}
C_{Ak}^{i}\rightarrow C_{AB}^{C},\text{ \ \ \ }C_{A_{1}\cdots
A_{p+1}}^{i}\rightarrow C_{A_{1}\cdots A_{p+1}}^{C}.
\end{equation}
To avoid confusion with the indices of the invariant tensor, we introduce a
comma to separate the indices corresponding to different sectors of the
algebra $g_{A_{1}\cdots A_{m}i_{1}\cdots i_{n}}\rightarrow g_{A_{1}\cdots
A_{m},B_{1}\cdots B_{n}}$. Note that the tensor is symmetric on the first
set of indices but it still can be symmetric or antisymmetric on the second
one depending on the value of $p$. In this case, the invariant tensor
conditions (\ref{it1}) - (\ref{it3}) take the form%
\begin{align}
\sum_{r=1}^{m}C_{A_{0}A_{r}}^{C}g_{A_{1}\cdots\hat{A}_{r}C\cdots
A_{m},B_{1}\cdots B_{n}}+\sum_{s=1}^{n}C_{A_{0}B_{s}}^{C}g_{A_{1}\cdots
A_{m},B_{1}\cdots\hat{B}_{s}C\cdots B_{n}} & =0,  \label{adj1} \\
\sum_{r=1}^{m+1}C_{A_{r}B_{1}\cdots B_{p}}^{C_{1}}g_{A_{1}\cdots\hat{A}%
_{r}\cdots A_{m+1}C_{1}\cdots C_{n}} & =0,  \label{adj2} \\
\sum_{r=1}^{m+1}C_{A_{0}A_{r}}^{C_{1}}g_{A_{1}\cdots\hat{A}_{r}\cdots
A_{m+1}C_{1}\cdots C_{n}} & =0.  \label{adj3}
\end{align}
Eq. (\ref{adj1}) is equivalent to the invariant tensor condition for Lie
algebras. This makes it easier to find invariant tensors for FDAs; when the $%
p$-form is in the adjoint representation of the Lie subalgebra, an invariant
tensor of the whole FDA is an invariant tensor of the Lie subalgebra that
also satisfies the second and third conditions (\ref{adj2}) and (\ref{adj3}).

\section{Chern--Weil theorem}

Transgression and CS forms are chosen as Lagrangians for gravitational
theories due to their invariance properties under gauge transformations. In
particular, transgression forms are completely invariant and allow finding
conserved charges. In the second order formalism of gravity, the fundamental
field is the spacetime metric $g_{\mu \nu }$ and not the Levi-Civita
connection $\Gamma _{\mu \nu }^{\lambda }$. The Levi--Civita connection is
present in the formulation, but it is completely determined by the metric
tensor. However, the first-order formalism allows one to consider both the
metric and the connection as off-shell independent fields, each encoded in
the components of one-forms evaluated in a Lie algebra. Since Chern--Simons
and transgression theories are background-free theories that depend on a
one-form gauge connection, they are good candidates to be considered
gravitational theories, generalizing general relativity and introducing
gauge invariance under a certain Lie group. See Refs. \cite{subs, tf} for detailed
reviews on the relation between invariant densities, CS and transgression
forms, and physical theories. Gravitational theories in which
the Lagrangians are CS forms and the invariance is described by space-time
Lie groups, namely Poincar\'{e} and (Anti) de-Sitter groups, were proposed
in Refs. \cite{nie, achu, wi1} for the three-dimensional case. In the last
decade of the 20th century, A. Chamseddine extended CS forms to higher
dimensions \cite{cham,cham2}. The supersymmetrization of such models was
introduced and extensively studied in Refs. \cite%
{ban,zan,tron,tron2,tron3,has}.

\subsection{CS and transgression forms}

The relation between CS and transgression forms emerges naturally in the
Chern--Weil theorem: Let $\mu$ and $\bar{\mu}$ be one-form gauge connections
on a $2n+1$ dimensional manifold, evaluated on a Lie algebra. Let $R=\mathrm{%
d}\mu+\mu^{2}$ and $\bar{R}=\mathrm{d}\bar{\mu}+\bar{\mu}^{2}$ their
corresponding two-form curvatures. Then%
\begin{equation}
\mathrm{Tr}\text{~}R^{n}-\mathrm{Tr}\text{~}\bar{R}^{n}=\mathrm{d}%
Q_{2n-1}\left( \mu,\bar{\mu}\right) ,
\end{equation}
where $\mathrm{Tr}$ denotes the trace over the elements of the Lie algebra
on which the curvatures are valuated. The $\left( 2n-1\right) $-form $%
Q_{2n-1}\left( \mu,\bar{\mu}\right) $ is called transgression form and its
explicit expression can be found by introducing a homotopic gauge field $%
\mu_{t}=\bar{\mu}+t\left( \mu-\bar{\mu}\right) $ with its corresponding
homotopic curvature $R_{t}=\mathrm{d}\mu_{t}+\mu_{t}^{2}$ depending on a
parameter $t\in\left[ 0,1\right] $. Note that the homotopic parameter
interpolates $\mu_{t}$ between $\bar{\mu}$ and $\mu$ and therefore, also $%
R_{t}$ between $\bar{R}$ and $R$. Then, it is possible to write down the
transgression $\left( 2n-1\right) $-form as%
\begin{equation}
Q_{2n-1}\left( \mu,\bar{\mu}\right) =n\int_{0}^{1}\mathrm{d}t~\mathrm{Tr}%
\text{~}\left[ \left( \mu-\bar{\mu}\right) R_{t}^{n-1}\right].
\end{equation}
By locally setting $\bar{\mu}=0$, the transgression form become a CS-form $%
Q_{2n-1}\left( \mu\right) =Q_{2n-1}\left( \mu,0\right) $, satisfying the
well-known relation%
\begin{equation}
\mathrm{Tr}\text{~}R^{n}=\mathrm{d}Q_{2n-1}\left( \mu\right) .
\end{equation}
Since the connection $\bar{\mu}$ cannot be globally fixed, the CS form can
be only locally well defined.

\subsection{Extended Chern--Weil theorem}

In order to generalize the Chern--Weil theorem to the case of FDAs, let us
consider the invariant density from Eq. (\ref{inv}) for the homotopic gauge
fields $\mu_{t}^{A}$ and $B_{t}^{i}$%
\begin{align}
\mu_{t}^{A} & =\left( 1-t\right) \mu_{0}^{A}+t\mu_{1}^{A}, \\
B_{t}^{i} & =\left( 1-t\right) B_{0}^{i}+tB_{1}^{i},
\end{align}
with $t\in\left[ 0,1\right] $. Since the homotopic parameter $t$
interpolates between $\left( \mu_{0},B_{0}\right) $ and $\left( \mu
_{1},B_{1}\right) $, the difference $\chi_{q}\left( \mu_{1},B_{1}\right)
-\chi_{q}\left( \mu_{0},B_{0}\right) $ can be written as the following
integral%
\begin{equation}
\chi_{q}\left( \mu_{1},B_{1}\right) -\chi_{q}\left( \mu_{0},B_{0}\right)
=\int_{0}^{1}\mathrm{d}t\frac{\mathrm{d}}{\mathrm{d}t}\chi_{q}\left( \mu
_{t},B_{t}\right) .  \label{xi10}
\end{equation}
Using the definition of covariant derivative for the FDA and performing the
derivation inside of the integral, it is possible to show that the
difference $\chi_{q}\left( \mu_{1},B_{1}\right) -\chi_{q}\left(
\mu_{0},B_{0}\right) $ is an exact form%
\begin{equation}
\chi_{q}\left( \mu_{1}\right) -\chi_{q}\left( \mu_{0}\right) =\mathrm{d}%
Q_{q-1}\left( \mu_{1},\mu_{0}\right) ,
\end{equation}
where the $\left( q-1\right) $-form $Q_{q-1}\left( \mu_{1},\mu_{0}\right) $
is explicitly given by%
\begin{align}
Q_{q-1}\left( \mu_{1},\mu_{0}\right) & =\sum_{m,n\in q\left( p\right)
}g_{A_{1}\cdots A_{m}i_{1}\cdots i_{n}}\int_{0}^{1}\mathrm{d}t\left(
mu^{A_{1}}R_{t}^{A_{2}}\cdots R_{t}^{A_{m}}R_{t}^{i_{1}}\cdots
R_{t}^{i_{n}}\right.  \notag \\
& \left. +nR_{t}^{A_{1}}\cdots R_{t}^{A_{m}}u^{i_{1}}R_{t}^{i_{2}}\cdots
R_{t}^{i_{n}}\right) .   \label{transgression}
\end{align}%
Here we define $u^{A}=\mu_{1}^{A}-\mu_{0}^{A}$ and $%
b^{i}=B_{1}^{i}-B_{0}^{i} $. $R_{t}^{A}$ and $R_{t}^{i}$ are the curvatures
corresponding to the homotopic gauge fields. For convenience, we rename $%
B^{i}=\mu^{i}$, $B_{t}^{i}=\mu_{t}^{i}$ and $b^{i}=u^{i}$. Then we can write 
$\mu=\left( \mu^{A},\mu^{i}\right) $ as a single composite field and the
same for the corresponding curvature. We can also write $\chi_{q}\left(
\mu_{t}^{A},B_{t}^{i}\right) =\chi_{q}\left( \mu_{t}^{A},\mu_{t}^{i}\right)
=\chi _{q}\left( \mu_{t}\right) $. This allows to write the generalization
of the Chern--Weil theorem in a compact way that will be useful in future
calculations%
\begin{equation}
\chi_{q}\left( \mu_{1}\right) -\chi_{q}\left( \mu_{0}\right) =\mathrm{d}%
Q_{q-1}\left( \mu_{1},\mu_{0}\right) .
\end{equation}
As with the standard transgression forms, if we fix $\mu_{0}=\left( \mu
_{0}^{A},\mu_{0}^{i}\right) =\left( 0,0\right) $ in Eq. (\ref{transgression}%
), we obtain the generalization of the Chern--Simons form to the case of
FDAs. Analogously to the case with Lie algebras, this can be done only
locally.

\section{Gauge anomalies}

The presence of anomalies in a theory is due to the breaking of classical
symmetries in the quantization process. The chiral anomaly, introduced in
Refs \cite{schw,rosem,adler,bard} appears in gauge theories that interact
with Weyl fermions. The $U_{A}(1)$ or abelian anomaly is given by the
divergence of the classically conserved current, and it is proportional to
the Chern--Pontryagin $4$-form%
\begin{equation}
\partial _{\mu }J^{\mu }\propto \mathrm{Tr~}\left[ \varepsilon _{\mu \nu
\rho \sigma }R^{\mu \nu }R^{\rho \sigma }\right] =\mathrm{Tr~}\left[
\partial ^{\mu }\varepsilon _{\mu \nu \rho \sigma }\left( \mu ^{\nu
}\partial ^{\rho }\mu ^{\sigma }+\frac{2}{3}\mu ^{\nu }\mu ^{\rho }\mu
^{\sigma }\right) \right] ,  \label{div}
\end{equation}%
where $\mu _{\nu }=\mu _{\nu }^{A}T_{A}$ is a gauge connection valuated on
the Lie algebra of an internal Lie group with generators $t_{A}$ \cite{zumnb}%
. By introducing the one-form $\mu =\mu _{\mu }^{A}$d$x^{\mu }t_{A}$ and the
exterior derivative $\mathrm{d}=\mathrm{d}x^{\mu }\partial _{\mu }$ it is
possible to write the divergence (\ref{div}) in terms of differential forms
and the Hodge operator%
\begin{equation}
\mathrm{d}\ast J\propto \mathrm{Tr~}R^{2}=\mathrm{d}Q_{3}\left( \mu \right) =%
\mathrm{Tr~}\left[ \mathrm{d}\left( \mu \mathrm{d}\mu +\frac{2}{3}\mu
^{3}\right) \right] ,  \label{dq}
\end{equation}%
where $Q_{3}\left( \mu \right) $ is the standard CS $3$-form.

On the other side, the so called non-abelian anomaly is given by the
covariant divergence of a non-abelian current $J_{\mu A}$%
\begin{equation}
\mathrm{D}^{\mu }J_{\mu A}\propto \mathrm{Tr~}\partial ^{\mu }\left[
t_{A}\varepsilon _{\mu \nu \rho \sigma }\left( \mu ^{\nu }\partial ^{\rho
}\mu ^{\sigma }+\frac{1}{2}\mu ^{\nu }\mu ^{\rho }\mu ^{\sigma }\right) %
\right] ,
\end{equation}%
or, in terms of differential forms%
\begin{equation}
\mathrm{D}\ast J_{A}\propto \mathrm{d}\left[ \mathrm{Tr~}\left( t_{A}\left(
\mu \mathrm{d}\mu +\frac{1}{2}\mu ^{3}\right) \right) \right] .  \label{dj}
\end{equation}
In general, it is not possible to write the right-hand side of Eqs. (\ref{dj}%
) in terms of gauge fields and field strengths. However, an interesting
result due to B. Zumino in Refs. \cite{zumnb,zumnc} shows that the
non-abelian gauge anomaly can be derived from the gauge-variation of the CS
form on Eq. (\ref{dq}), which is porportional to the exterior derivative of
certain two-form $Q_{2}^{1}\left( \varepsilon ,\mu \right) $ depending on
the gauge fields and the $0$-form parameter of the transformation $%
\varepsilon =\varepsilon ^{A}T_{A}$%
\begin{equation}
\delta Q_{3}\left( \mu \right) =\mathrm{d}Q_{2}^{1}\left( \varepsilon ,\mu
\right) .
\end{equation}%
Those results are also valid in $2n$-dimensional spacetimes, being in
general possible to write down the abelian anomaly in terms of the total
derivative of a $\left( 2n-1\right) $-dimensional CS form%
\begin{equation}
\mathrm{d}\ast J\propto \mathrm{Tr~}\left( R^{n}\right) =\mathrm{d}%
Q_{2n-1}\left( \mu \right) .  \label{ab}
\end{equation}%
In the same way, the non-abelian anomaly in a $\left( 2n-2\right) $%
-dimensional spacetime can derived from the $2n$-dimensional abelian anomaly
on Eq. (\ref{ab}) by introducing a set of Lorentz-scalar fields $%
\varepsilon ^{A}$ and computing the gauge variation of the $\left( 2n-1\right) $%
-dimensional CS form 
\begin{equation}
\delta Q_{2n-1}\left( \mu \right) =\mathrm{d}Q_{2n-2}^{1}\left( \varepsilon
,\mu \right) .
\end{equation}%
The $\left( 2n-2\right) $-form $Q_{2n-2}^{1}\left( \varepsilon ,\mu \right) $
has the following integral representation \cite{zumnb,zumnc}%
\begin{equation}
Q_{2n-2}^{1}\left( \varepsilon ,\mu \right) =n(n-1)\int_{0}^{1}\mathrm{d}%
t(1-t)\mathrm{Str}\left( \varepsilon ,\mathrm{d}\left( \mu
,R_{t}^{n-2}\right) \right) ,  \label{chiral}
\end{equation}%
where $\mathrm{Str}$ denotes the symmetrized trace over the algebraic
elements.

\subsection{Extended anomalies}

Recently, it was found that it is possible to find gauge-invariant densities
in the context of higher gauge theory\footnote{%
See for instance Refs. \cite{sav5,savv1,savv2,savv3,snpb}}. Moreover, in
Ref. \cite{epjs,sav4} it was studied the existence of gauge anomalies
generated by such invariants. In this section we study the existence of
anomalies, starting from the extended CS forms introduced in previous
sections. Let us consider the following gauge field%
\begin{equation}
\mu=\left( \mu^{A},\mu^{i}\right) ,
\end{equation}
consisting on one-forms $\mu^{A}$ and $p$-forms $\mu^{i}$. The
corresponding curvature is given by a $\left( 2,p+1\right) $-form $R=\left(
R^{A},R^{i}\right) $. As we have seen, a general field $X$ can be decomposed
as a set $X=\left( X^{A},X^{i}\right) $, where $X^{A}$ is a $r$-form and $%
X^{i}$ is a $\bar{r}$-form with $\bar {r}=r+p-1$. In such case we say that
the array $X$ is of degree $r$. Let $X$ be an array of degree $x$ and $Y$ be
an array of degree $y$. We introduce the following products between two
arrays%
\begin{equation}
Z=\left( Z^{A},Z^{i}\right) =\left[ X,Y\right] ,
\end{equation}
where the components of $Z$ are given by 
\begin{align}
Z^{A} & =\left[ X,Y\right] ^{A}=C_{BC}^{A}X^{B}Y^{C},  \label{z1} \\
Z^{i} & =\left[ X,Y\right] ^{i}=C_{Bj}^{i}X^{B}Y^{j}.  \label{z2}
\end{align}
Note that Eqs. (\ref{z1}) and (\ref{z2}) imply that $Z^{A}$ is a $\left(
x+y\right) $-form and $Z^{i}$ is a $\left( x+y+p-1\right) $-form.

Let us consider now the arrays $X_{r}\ $and $Y_{s}$ of degree $x_{r}$ and $%
y_{s}$ respectively. We introduce the product between $p+1$ arrays with the
following components%
\begin{align}
\left[ X_{1},\ldots,X_{p+1}\right] ^{A} & =0, \\
\left[ X_{1},\ldots,X_{p+1}\right] ^{i} & =C_{A_{1}\cdots
A_{p+1}}^{i}X_{1}^{A_{1}}\cdots X_{p+1}^{A_{p+1}}.
\end{align}
On the other hand, as we will see later, it is convenient to introduce a
compact notation for the FDA invariant tensor. We denote 
\begin{equation}
\left\langle X_{1},\ldots,X_{m};Y_{1},\ldots,Y_{n}\right\rangle
=g_{A_{1}\cdots A_{m}i_{1}\cdots i_{n}}X_{1}^{A_{1}}\cdots
X_{m}^{A_{m}}Y_{1}^{i_{1}}\cdots Y_{n}^{i_{n}}.
\end{equation}
This bracket separates the element before and after the semicolon, being the
first ones evaluated in the Lie subalgebra and the latter on the extended
sector. It is easy to check the following (anti)symmetry rules%
\begin{align}
\left\langle \ldots,X_{r},X_{r+1},\ldots;Y_{1},\ldots,Y_{n}\right\rangle &
=\left( -1\right) ^{x_{r}x_{r+1}}\left\langle
\ldots,X_{r+1},X_{r},\ldots;Y_{1},\ldots,Y_{n}\right\rangle, \\
\left\langle X_{1},\ldots,X_{m};\ldots,Y_{s},Y_{s+1},\ldots\right\rangle &
=\left( -1\right) ^{\bar{y}_{s}\bar{y}_{s+1}+p+1}\left\langle
X_{1},\ldots,X_{m};\ldots,Y_{s+1},Y_{s},\ldots\right\rangle.
\end{align}
Now we recall the invariant tensor conditions (\ref{it1}) - (\ref{it3}).
Multiplying the first invariant tensor condition (\ref{it1}) by $\Theta
^{A_{0}}X^{A_{1}}\cdots X^{A_{m}}Y^{i_{1}}\cdots Y^{i_{n}}$ where $%
\Theta^{A} $ is the first component of the degree-$\theta$ array $%
\Theta=\left( \Theta ^{A},\Theta^{i}\right) $ we get 
\begin{dmath}
\sum_{r=1}^{m}\Theta^{A_{0}}X_{1}^{A_{1}}\cdots
X_{m}^{A_{m}}Y_{1}^{i_{1}%
}\cdots Y_{n}^{i_{n}}C_{A_{0}A_{r}}^{C}%
g_{A_{1}\cdots\hat{A}_{r}C\cdots A_{m}i_{1}\cdots i_{n}}\nonumber\\
+\sum_{s=1}^{n}\Theta^{A_{0}}X_{1}^{A_{1}}\cdots
X_{m}^{A_{m}}Y_{1}^{i_{1}%
}\cdots Y_{n}^{i_{n}}C_{A_{0}i_{s}}^{k}%
g_{A_{1}\cdots A_{m}i_{1}\cdots\hat{\imath}_{s}k\cdots i_{n}}=0.
\end{dmath}Using the new notation for the invariant tensor and the product
given in Eqs. (\ref{z1}) and (\ref{z2}) we obtain the following identity 
\begin{dmath}
\sum_{r=1}^{m}\left(  -1\right)  ^{\theta\left(  x_{1}+\cdots
+x_{r-1}\right)  }\left\langle X_{1},\ldots,X_{r-1},\left[  \Theta
,X_{r}\right]  ,X_{r+1},\ldots,X_{m};Y_{1},\ldots,Y_{n}\right\rangle
+\sum_{s=1}^{n}\left(  -1\right)  ^{\theta\left(  x_{1}+\cdots+x_{m}+\bar
{y}_{1}+\cdots\bar{y}_{s-1}\right)  }\left\langle X_{1},\ldots,X_{m}%
;Y_{1},\ldots,Y_{s-1},\left[  \Theta,Y_{s}\right]  ,Y_{s+1},\ldots
,Y_{n}\right\rangle    =0. \label{zum1}%
\end{dmath}In the same way, introducing a new set of arbitrary arrays $%
\Theta_{1},\ldots,\Theta_{p}$ and multiplying (\ref{it2}) and (\ref{it3}) by 
$\Theta^{B_{1}}\cdots\Theta^{B_{p}}X^{A_{1}}\cdots
X^{A_{m+1}}Y^{i_{2}}\cdots Y^{i_{n}}$ and $\Theta^{j}X^{A_{1}}\cdots
X^{A_{m+1}}Y^{i_{2}}\cdots Y^{i_{n}}$ respectively we obtain the second and
third invariant tensor conditions in terms of the new notation

\begin{align}
\sum_{r=1}^{m+1}\left( -1\right) ^{x_{r}\left( x_{r+1}+\cdots
+x_{m+1}\right) }\left\langle X_{1},\ldots,X_{r-1},X_{r+1},\ldots ,X_{m+1}; 
\left[ X_{r},\Theta_{1},\ldots,\Theta_{p}\right] ,Y_{2},\ldots,Y_{n}\right%
\rangle & =0.  \label{zum2} \\
\sum_{r=1}^{m+1}\left( -1\right) ^{x_{r}\left( x_{r+1}+\cdots
+x_{m+1}\right) }\left\langle X_{1},\ldots,X_{r-1},X_{r+1},\ldots ,X_{m+1}; 
\left[ X_{r},\Theta\right] ,Y_{2},\ldots,Y_{n}\right\rangle & =0.
\label{zum3}
\end{align}
Eqs. (\ref{zum1}) - (\ref{zum3}) are the generalization of the invariant
tensor property for Lie algebras, given in Eq. (B.10) from Ref. \cite{zumnb}.

With the new notations, the Chern--Weil theorem for CS forms can be written
in a more compact manner%
\begin{equation}
\chi_{q}\left( \mu\right) =\mathrm{d}Q_{q-1}\left( \mu\right) ,
\end{equation}
where%
\begin{equation}
Q_{q-1}\left( \mu\right) =\sum_{m,n\in q\left( p\right) }\int_{0}^{1}\mathrm{%
d}t\left( m\left\langle \mu,R_{t}^{m-1};R_{t}^{n}\right\rangle
+n\left\langle R_{t}^{m};\mu,R_{t}^{n-1}\right\rangle \right) .  \label{cs}
\end{equation}
The purpose of this notation is to be able to find a gauge anomaly from the
extended CS form in a more compact way. To achieve this, it is necessary to
find an expression for the gauge variation of $Q^{\left( q-1\right) }\left(
\mu\right) $ in terms of an exact form as with Lie groups 
\begin{equation}
\delta Q_{q-1}\left( \mu\right) =\mathrm{d}\omega_{q-2}^{1}\left(
\varepsilon,\mu\right) .
\end{equation}
It is also convenient to separate the independent variations with respect to
the parameters $\varepsilon^{A}$ and $\varepsilon^{i}$. Since the standard
and extended transformations are independent, this leaves to two different
generalizations of the gauge anomaly from Eq. (\ref{chiral}). Let us begin
with the extended variation, i.e., the one proportional to $\varepsilon^{i}$.

\subsection{Extended variations}

The gauge variations of the gauge fields and curvatures parametrized by the $%
\left( p-1\right) $-form $\varepsilon^{i}$ are given by%
\begin{align}
\delta\mu^{A} & =0,\text{ \ \ \ \ \ }\delta\mu^{i}=\mathrm{d}\varepsilon
^{i}+\left[ \mu,\varepsilon\right] ^{i}, \\
\delta R^{A} & =0,\text{ \ \ \ \ \ }\delta R^{i}=\left[ R,\varepsilon \right]
^{i}.
\end{align}
Using the generalized Jacobi identity, it is direct to prove that the
variation of the corresponding homotopic curvatures are%
\begin{equation}
\delta R_{t}^{A}=0,\text{ \ \ \ }\delta R_{t}^{i}=\left[ R_{t},\varepsilon %
\right] ^{i}+t\left( t-1\right) \left[ \mu,\mathrm{d}\varepsilon\right] ^{i}.
\end{equation}
The variation of the CS form is then given by%
\begin{align}
\delta Q_{q-1}\left( \mu\right) & =\sum_{m,n}\int_{0}^{1}\mathrm{d}t\left[
nm\left\langle \mu,R_{t}^{m-1};\left[ R_{t},\varepsilon\right]
,R_{t}^{n-1}\right\rangle +n\left\langle R_{t}^{m};\left[ \mu,\varepsilon %
\right] ,R_{t}^{n-1}\right\rangle +n\left\langle R_{t}^{m};\mathrm{d}%
\varepsilon,R_{t}^{n-1}\right\rangle \right.  \notag \\
& +t\left( t-1\right) mn\left\langle \mu,R_{t}^{m-1};\left[ \mu ,\mathrm{d}%
\varepsilon\right] ,R_{t}^{n-1}\right\rangle +n\left( n-1\right)
\left\langle R_{t}^{m};\mu,\left[ R_{t},\varepsilon\right]
,R_{t}^{n-2}\right\rangle  \notag \\
& +\left. t\left( t-1\right) n\left( n-1\right) \left\langle R_{t}^{m};\mu, 
\left[ \mu,\mathrm{d}\varepsilon\right] ,R_{t}^{n-2}\right\rangle \right] .
\end{align}
From Eqs. (\ref{zum1}) - (\ref{zum3}) it can be shown that the following
relations hold%
\begin{equation}
\left\langle R_{t}^{m};\left[ R_{t},\varepsilon\right] ,\mu,R_{t}^{n-2}%
\right\rangle =0,
\end{equation}%
\begin{equation}
\left\langle R_{t}^{m};\left[ \mu,\varepsilon\right] ,R_{t}^{n-1}\right%
\rangle +m\left\langle \mu,R_{t}^{m-1};\left[ R_{t},\varepsilon\right]
,R_{t}^{n-1}\right\rangle =0,
\end{equation}
\begin{dmath}
m\left\langle \left[  \mu,R_{t}\right]  ,R_{t}^{m-1};\mathrm{d}\varepsilon
,\mu,R_{t}^{n-2}\right\rangle +\left\langle R_{t}^{m};\left[  \mu
,\mathrm{d}\varepsilon\right]  ,\mu,R_{t}^{n-2}\right\rangle +\left(
-1\right)  ^{p}\left\langle R_{t}^{m};\mathrm{d}\varepsilon,\left[  \mu
,\mu\right]  ,R_{t}^{n-2}\right\rangle \nonumber
+\left(  n-2\right)  \left\langle R_{t}^{m};\mathrm{d}\varepsilon,\mu,\left[
\mu,R_{t}\right]  ,R_{t}^{n-3}\right\rangle =0,
\end{dmath}%
\begin{dmath}
\left\langle \left[  \mu,\mu\right]  ,R_{t}^{m-1};\mathrm{d}\varepsilon
,R_{t}^{n-1}\right\rangle -\left(  m-1\right)  \left\langle \mu,\left[
\mu,R_{t}\right]  ,R_{t}^{m-2};\mathrm{d}\varepsilon,R_{t}^{n-1}\right\rangle
-\left\langle \mu,R_{t}^{m-1};\left[  \mu,\mathrm{d}\varepsilon\right]
,R_{t}^{n-1}\right\rangle \nonumber
+\left(  -1\right)  ^{p+1}\left(  n-1\right)  \left\langle \mu,R_{t}%
^{m-1};\mathrm{d}\varepsilon,\left[  \mu,R_{t}\right]  ,R_{t}^{n-2}%
\right\rangle =0.
\end{dmath}This allows to remove the terms including brackets between $\mu$
and d$\varepsilon$ and write the gauge variation as follows%
\begin{dmath}
\delta Q_{q-1}\left(  \mu\right)     =\sum_{m,n}\int_{0}^{1}\mathrm{d}%
t~t\left(  t-1\right)  \left[  mn\left\langle \left[  \mu,\mu\right]
,R_{t}^{m-1};\mathrm{d}\varepsilon,R_{t}^{n-1}\right\rangle \right.
-mn\left(  m-1\right)  \left\langle \mu,\left[  \mu,R_{t}\right]  ,R_{t}%
^{m-2};\mathrm{d}\varepsilon,R_{t}^{n-1}\right\rangle \\
+mn\left(  n-1\right)  \left(  -1\right)  ^{p+1}\left\langle \mu,R_{t}%
^{m-1};\mathrm{d}\varepsilon,\left[  \mu,R_{t}\right]  ,R_{t}^{n-2}%
\right\rangle +n\left\langle R_{t}^{m};\mathrm{d}\varepsilon,R_{t}%
^{n-1}\right\rangle \\
+n\left(  n-1\right)  m\left(  -1\right)  ^{p}\left\langle \left[  \mu
,R_{t}\right]  ,R_{t}^{m-1};\mathrm{d}\varepsilon,\mu,R_{t}^{n-2}\right\rangle
+n\left(  n-1\right)  \left\langle R_{t}^{m};\mathrm{d}\varepsilon,\left[
\mu,\mu\right]  ,R_{t}^{n-2}\right\rangle \\
+\left.  n\left(  n-1\right)  \left(  n-2\right)  \left(  -1\right)
^{p}\left\langle R_{t}^{m};\mathrm{d}\varepsilon,\mu,\left[  \mu,R_{t}\right]
,R_{t}^{n-3}\right\rangle \right].
\end{dmath}The next step is to use the generalized Bianchi identities, the
definition of the homotopic curvatures and Eqs. (\ref{zum1}) -\ (\ref{zum3})
to obtain the following relations%
\begin{align}
\left[ \mu_{t},R_{t}\right] ^{A} & =-\mathrm{d}R_{t}^{A},  \label{rel1} \\
\left[ \mu_{t},R_{t}\right] ^{i} & =-\mathrm{d}R_{t}^{i}+\left[ R_{t},\mu_{t}%
\right] ^{i}+\frac{1}{p!}\left[ R_{t},\mu_{t}^{p}\right] ^{i},
\end{align}%
\begin{align}
t\left[ \mu,\mu\right] ^{A} & =\frac{\partial R_{t}^{A}}{\partial t}-\mathrm{%
d}\mu^{A}, \\
t\left[ \mu,\mu\right] ^{i} & =\frac{1}{2}\frac{\partial R_{t}^{i}}{\partial
t}-\frac{1}{2}\mathrm{d}\mu^{i}-\frac{t^{p}}{2p!}\left[ \mu ^{p+1}\right]
^{i},
\end{align}%
\begin{align}
\left\langle R_{t}^{m};\left[ \mu,\mu_{t}\right] ,\mathrm{d}\varepsilon
,R_{t}^{n-2}\right\rangle -m\left\langle \mu,R_{t}^{m-1};\mathrm{d}%
\varepsilon,\left[ R_{t},\mu_{t}\right] ,R_{t}^{n-2}\right\rangle & =0, \\
\left\langle R_{t}^{m};\left[ \mu,\mu_{t}\right] ,\mathrm{d}\varepsilon
,R_{t}^{n-2}\right\rangle +m\left\langle \mu,R_{t}^{m-1};\left[ R_{t},\mu
_{t}\right] ,\mathrm{d}\varepsilon,R_{t}^{n-2}\right\rangle & =0.
\label{rel6}
\end{align}
We use relations (\ref{rel1}) -\ (\ref{rel6}) to write the extended
variation of $Q^{\left( q-1\right) }\left( \mu\right) $ in terms of the
total derivatives $\partial/\partial t$ and d 
\begin{align}
\delta Q_{q-1}\left( \mu\right) & =\sum_{m,n}\int_{0}^{1}\mathrm{d}t\left[ n%
\frac{\partial}{\partial t}\left\{ \left( t-1\right) \left\langle R_{t}^{m};%
\mathrm{d}\varepsilon,R_{t}^{n-1}\right\rangle \right\} \right.  \notag \\
& -\left( t-1\right) mn\left( -1\right) ^{\left( p+1\right) \left(
n-1\right) }\left\langle \mathrm{d}\left( \mu,R_{t}^{m-1};R_{t}^{n-1}\right)
,\mathrm{d}\varepsilon\right\rangle  \notag \\
& -\left. \left( t-1\right) \left( -1\right) ^{\left( p+1\right) \left(
n-1\right) }n\left( n-1\right) \left\langle \mathrm{d}\left(
R_{t}^{m};\mu,R_{t}^{n-2}\right) ,\mathrm{d}\varepsilon\right\rangle \right]
.
\end{align}
Note that the first term vanishes while the second and third terms are exact
forms. This means that we can write the gauge variation of the Chern-Simons
form (\ref{cs}) in terms of a $\left( q-2\right) $-form proportional to $%
\varepsilon^{i}$ which generalizes the expression for the non-abelian gauge
anomaly%
\begin{equation}
\delta_{\text{Extended}}Q_{q-1}\left( \mu\right) =\mathrm{d}%
\omega_{q-2}^{1}\left( \varepsilon^{i},\mu\right) ,
\end{equation}
where%
\begin{equation}
\omega_{q-2}^{1}\left( \varepsilon^{i},\mu\right) =\sum_{m,n}\int_{0}^{1}%
\mathrm{d}t\left( 1-t\right) n\left( m\left\langle \mathrm{d}\left(
\mu,R_{t}^{m-1};R_{t}^{n-1}\right) ,\varepsilon\right\rangle +\left(
n-1\right) \left\langle \mathrm{d}\left( R_{t}^{m};\mu,R_{t}^{n-2}\right)
,\varepsilon\right\rangle \right) .  \label{anom1}
\end{equation}

\subsection{Standard variations}

Let us consider now the standard variations of gauge fields and curvatures.
This case presents some differences because, in order to obtain the anomaly
term, we need to write the CS form in terms of new homotopic curvatures. Let
us introduce some definitions before we proceed on this.

Given a degree-$1$ array $\mu=\left( \mu^{A},\mu^{i}\right) $ of gauge
fields, we have a corresponding degree-$2$ array of curvatures $R=\left(
R^{A},R^{i}\right) $. In terms of $\mu$ and $R$ we define the derivative
operator \cite{zumnb}%
\begin{align}
\mathrm{d}\mu^{A} & =R^{A}-\frac{1}{2}\left[ \mu,\mu\right] ^{A}, \\
\mathrm{d}\mu^{i} & =R^{i}-\left[ \mu,\mu\right] ^{i}-\frac{1}{\left(
p+1\right) !}\left[ \mu^{p+1}\right]^{i}.
\end{align}
Using the Jacobi identities we can check that this definition verifies the
nilpotent condition $\mathrm{d}^{2}=0$.

Now we consider an arbitrary variation on $\mu$ and $R$. With respect to
that variation we introduce the homotopy operator $\ell$ such that its
action on the arrays $\mu$ and $R$ is given by%
\begin{equation}
\ell\mu=0,\text{ \ \ \ }\ell R=\delta\mu.
\end{equation}
By direct inspection it can be shown that the homotopy operator $\ell$
satisfy the following anticommuting relations%
\begin{equation}
\left( \ell\mathrm{d}+\mathrm{d}\ell\right) \mu=\delta\mu\text{, \ \ }\left(
\ell\mathrm{d}+\mathrm{d}\ell\right) R=\delta R.
\end{equation}
Now we introduce new homotopic gauge fields. The one-form $\mu^{A}$ is
parametrized as usual. However, the $p$-form $\mu^{i}$ is parametrized with
a power of the parameter, such that $\mu_{t}$ still interpolates between $%
\mu_{0}$ and $\mu_{1}$ along a convenient trajectory in the parametric space 
\begin{equation}
\mu_{t}^{A}=t\mu^{A},\text{ \ \ }\mu_{t}^{i}=t^{p}\mu^{i}.  \label{homotopy}
\end{equation}
If we consider a variation along the parameter $t$, then it is possible to
define a homotopic operator $\ell_{t}$ with respect to such variation
satisfying the relation $\ell\mathrm{d}+\mathrm{d}\ell=\mathrm{d}_{t}$%
\begin{equation}
\ell_{t}\mu_{t}=0,\text{ \ \ \ \ }\ell_{t}R_{t}=\mathrm{d}_{t}\mu _{t}=%
\mathrm{d}t\frac{\partial\mu_{t}}{\partial t}.
\end{equation}
Integrating the homotopic operator between $0$ and $1$ we find 
\begin{equation}
\int_{0}^{1}\left( \ell_{t}\mathrm{d}\text{+}\mathrm{d}\ell_{t}\right)
=\int_{0}^{1}\mathrm{d}_{t}.  \label{ld}
\end{equation}
By applying the left-hand side of (\ref{ld}) into $\chi_{q}\left( \mu
_{t}\right) $ we use Stokes' theorem and recover in this way the Chern-Weil
theorem%
\begin{equation}
\int_{0}^{1}\left( \ell_{t}\mathrm{d}\text{+}\mathrm{d}\ell_{t}\right)
\chi_{q}\left( \mu_{t}\right) =\int_{0}^{1}\mathrm{d}t\frac{\partial }{%
\partial t}\chi_{q}\left( \mu_{t}\right) =\chi_{q}\left( \mu\right) ,
\end{equation}
where%
\begin{equation}
\chi_{q}\left( \mu\right) =\mathrm{d}\int_{0}^{1}\ell_{t}\chi_{q}\left(
\mu_{t}\right) =\mathrm{d}Q_{q-1}\left( \mu\right) .
\end{equation}
A direct calculation of the action of the homotopy operator on $\chi
_{q}\left( \mu_{t}\right) $ leads to an explicit expression of the CS form.
Then we can write the generalized CS form in terms of the new homotopy (\ref%
{homotopy}) 
\begin{align}
Q_{q-1}\left( \mu\right) & =\int_{0}^{1}\ell_{t}\sum_{m,n}\left\langle
R_{t}^{m};R_{t}^{n}\right\rangle  \notag \\
& =\sum_{m,n}\int_{0}^{1}\mathrm{d}t\left( m\left\langle
\mu,R_{t}^{m-1};R_{t}^{n}\right\rangle +npt^{p-1}\left\langle
R_{t}^{m};\mu,R_{t}^{n-1}\right\rangle \right) .
\end{align}
This expression is equivalent to Eq. (\ref{cs}), the difference being that
it is written in terms of the new homotopy curvatures. The homotopic
curvature $R_{t}^{A}$ remains the same while the extended one, as $\mu_{t}$,
has a different dependency on $t$. Since the final expression is independent
of the integration, it is natural to have different choices.\ The new
expression must be chosen such that the anomaly is written in a more
convenient way. In this case, the variation of the CS-form is given by%
\begin{align}
\delta Q_{q-1}\left( \mu\right) & =\sum_{m,n}\int_{0}^{1}\mathrm{d}t\left[
m\left\langle \delta\mu,R_{t}^{m-1};R_{t}^{n}\right\rangle +m\left(
m-1\right) \left\langle \mu,\delta R_{t},R_{t}^{m-2};R_{t}^{n}\right\rangle
\right.  \notag \\
& +mn\left\langle \mu,R_{t}^{m-1};\delta R_{t},R_{t}^{n-1}\right\rangle
+mnpt^{p-1}\left\langle \delta R_{t},R_{t}^{m-1};\mu,R_{t}^{n-1}\right\rangle
\notag \\
& +\left. npt^{p-1}\left\langle R_{t}^{m};\delta\mu,R_{t}^{n-1}\right\rangle
+n\left( n-1\right) pt^{p-1}\left\langle R_{t}^{m};\mu,\delta
R_{t},R_{t}^{n-2}\right\rangle \right] .  \label{delta}
\end{align}
The $\varepsilon^{A}$-gauge variation of the gauge fields and homotopic
gauge curvatures is then given by

\begin{align}
\delta\mu^{A} & =\mathrm{d}\varepsilon^{A}+\left[ \mu,\varepsilon\right]
^{A}, \\
\delta\mu^{i} & =-\left[ \varepsilon,\mu_{t}\right] ^{i}-\frac{1}{p!}\left[
\varepsilon,\mu^{p}\right] ^{i}, \\
\delta R_{t}^{A} & =\left[ R_{t},\varepsilon\right] ^{A}+\left(
t^{2}-t\right) \left[ \mathrm{d}\varepsilon,\mu\right] ^{A},
\end{align}%
\begin{equation}
\delta R_{t}^{i}=-\left[ \varepsilon,R_{t}\right] ^{i}-\frac{t^{p-1}}{\left(
p-1\right) !}\left[ \varepsilon,R_{t},\mu^{p-1}\right] ^{i}+t^{p}\left(
t-1\right) \left[ \mathrm{d}\varepsilon,\mu\right] ^{i}+\frac{t^{p}\left(
t-1\right) }{p!}\left[ \mathrm{d}\varepsilon,\mu ^{p}\right] ^{i}.
\end{equation}
A different choice of the homotopy leads to a more complicated expression
for $\delta R_{t}^{i}$. This problem is not present in the previous case,
and therefore, we only introduce the new homotopy rule for the anomaly
resulting from the standard variations.

One more time we need to isolate the total derivative in Eq. (\ref{delta}).
To archieve this, we use Eqs. (\ref{zum1}) - (\ref{zum3}) prove the
following relations 
\begin{dmath}
\left\langle \left[  \mu,\mu\right]  ,\mathrm{d}\varepsilon,R_{t}^{m-2}%
;R_{t}^{n}\right\rangle -\left\langle \mu,\left[  \mathrm{d}\varepsilon
,\mu\right]  ,R_{t}^{m-2};R_{t}^{n}\right\rangle +\left(  m-2\right)
\left\langle \mu,\mathrm{d}\varepsilon,\left[  \mu,R_{t}\right]  ,R_{t}%
^{m-3};R_{t}^{n}\right\rangle
+n\left\langle \mu,\mathrm{d}\varepsilon,R_{t}^{m-2};\left[  \mu,R_{t}\right]
,R_{t}^{n-1}\right\rangle =0,
\end{dmath}

\begin{dmath}
\left\langle \left[  \mathrm{d}\varepsilon,\mu\right]  ,R_{t}^{m-1};\mu
,R_{t}^{n-1}\right\rangle -\left(  m-1\right)  \left\langle \mathrm{d}%
\varepsilon,\left[  \mu,R_{t}\right]  ,R_{t}^{m-2};\mu,R_{t}^{n-1}%
\right\rangle -\left\langle \mathrm{d}\varepsilon,R_{t}^{m-1};\left[  \mu
,\mu\right]  ,R_{t}^{n-1}\right\rangle -\left(  -1\right)  ^{p}\left(  n-1\right)  \left\langle \mathrm{d}%
\varepsilon,R_{t}^{m-1};\mu,\left[  \mu,R_{t}\right]  ,R_{t}^{n-2}%
\right\rangle =0,
\end{dmath}

\begin{dmath}
\left\langle \mu,R_{t}^{m-1};\left[  \mathrm{d}\varepsilon,\mu^{p}\right]
,R_{t}^{n-1}\right\rangle -\left\langle \mathrm{d}\varepsilon,R_{t}%
^{m-1};\left[  \mu^{p+1}\right]  ,R_{t}^{n-1}\right\rangle \\-\left(m-1\right)  \left\langle \mathrm{d}\varepsilon,\mu,R_{t}^{m-2};\left[
R_{t},\mu^{p}\right]  ,R_{t}^{n-1}\right\rangle =0,
\end{dmath}

\begin{dmath}
\left\langle \mu,R_{t}^{m-1};\left[  \mathrm{d}\varepsilon,\mu\right]
,R_{t}^{n-1}\right\rangle -\left\langle \mathrm{d}\varepsilon,R_{t}%
^{m-1};\left[  \mu,\mu\right]  ,R_{t}^{n-1}\right\rangle -\left(  m-1\right)
\left\langle \mathrm{d}\varepsilon,\mu,R_{t}^{m-2};\left[  R_{t},\mu\right]
,R_{t}^{n-1}\right\rangle =0,
\end{dmath}

\begin{dmath}
\left\langle R_{t}^{m};\mu,\left[  \mathrm{d}\varepsilon,\mu\right]
,R_{t}^{n-2}\right\rangle -\left(  -1\right)  ^{p}m\left\langle \mathrm{d}%
\varepsilon,R_{t}^{m-1};\left[  R_{t},\mu\right]  ,\mu,R_{t}^{n-2}%
\right\rangle =0,
\end{dmath}

\begin{dmath}
\left\langle R_{t}^{m};\left[  \mathrm{d}\varepsilon,\mu^{p}\right]
,\mu,R_{t}^{n-2}\right\rangle +m\left\langle \mathrm{d}\varepsilon,R_{t}%
^{m-1};\left[  R_{t},\mu^{p}\right]  ,\mu,R_{t}^{n-2}\right\rangle =0.
\end{dmath}This allows one to eliminate the terms with a bracket of the type 
$\left[ \mathrm{d}\varepsilon,\mu\right] $. The variation of the CS form
takes then the form%
\begin{align}
&  \delta Q_{q-1}\left(  \mu\right)  \nonumber\\
&  =\sum_{m,n}\int_{0}^{1}\mathrm{d}t~m\left[  \left\langle \mathrm{d}%
\varepsilon,R_{t}^{m-1};R_{t}^{n}\right\rangle +\left(  t-1\right)  \left(
\left(  m-1\right)  t\left(  m-2\right)  \left\langle \mu,\mathrm{d}%
\varepsilon,\left[  \mu,R_{t}\right]  ,R_{t}^{m-3};R_{t}^{n}\right\rangle
\right.  \right.  \nonumber\\
&  +\left(  m-1\right)  tn\left\langle \mu,\mathrm{d}\varepsilon,R_{t}%
^{m-2};\left[  \mu,R_{t}\right]  ,R_{t}^{n-1}\right\rangle +t^{p}n\left(
m-1\right)  \left\langle \mathrm{d}\varepsilon,\mu,R_{t}^{m-2};\left[
R_{t},\mu\right]  ,R_{t}^{n-1}\right\rangle \nonumber\\
&  +\frac{t^{p}}{p!}n\left(  m-1\right)  \left\langle \mathrm{d}%
\varepsilon,\mu,R_{t}^{m-2};\left[  R_{t},\mu^{p}\right]  ,R_{t}%
^{n-1}\right\rangle +n\left(  n-1\right)  pt^{2p-1}\left(  -1\right)
^{p}\left\langle \mathrm{d}\varepsilon,R_{t}^{m-1};\left[  R_{t},\mu\right]
,\mu,R_{t}^{n-2}\right\rangle \nonumber\\
&  +\frac{t^{p}}{p!}n\left(  n-1\right)  pt^{p-1}\left(  -1\right)
^{p}\left\langle \mathrm{d}\varepsilon,R_{t}^{m-1};\left[  R_{t},\mu
^{p}\right]  ,\mu,R_{t}^{n-2}\right\rangle +npt^{p-1}t\left(  m-1\right)
\left\langle \mathrm{d}\varepsilon,\left[  \mu,R_{t}\right]  ,R_{t}^{m-2}%
;\mu,R_{t}^{n-1}\right\rangle \nonumber\\
&  +\left(  m-1\right)  t\left\langle \left[  \mu,\mu\right]  ,\mathrm{d}%
\varepsilon,R_{t}^{m-2};R_{t}^{n}\right\rangle +t^{p}\left(  1+p\right)
n\left\langle \mathrm{d}\varepsilon,R_{t}^{m-1};\left\{  \left[  \mu
,\mu\right]  \right\}  ,R_{t}^{n-1}\right\rangle \nonumber\\
&  +\left.  \left.  \frac{t^{p}}{p!}n\left\langle \mathrm{d}\varepsilon
,R_{t}^{m-1};\left[  \mu^{p+1}\right]  ,R_{t}^{n-1}\right\rangle +\left(
-1\right)  ^{p}npt^{p-1}t\left(  n-1\right)  \left\langle \mathrm{d}%
\varepsilon,R_{t}^{m-1};\mu,\left[  \mu,R_{t}\right]  ,R_{t}^{n-2}%
\right\rangle \right)  \right]
\end{align}
Using one more time the definition of homotopic gauge curvatures and
generalized Bianchi identities we get the following identities%
\begin{align}
t\left[ \mu,\mu\right] ^{A} & =\frac{\partial R_{t}^{A}}{\partial t}-\mathrm{%
d}\mu^{A}, \\
\left( p+1\right) t^{p}\left[ \mu,\mu\right] ^{i}+\frac{t^{p}}{p!}\left[
\mu,\ldots,\mu\right] ^{i} & =\frac{\partial R_{t}^{i}}{\partial t}-pt^{p-1}%
\mathrm{d}\mu^{i}, \\
\left[ \mu_{t},R_{t}\right] ^{A} & =-\mathrm{d}R_{t}^{A}, \\
\left[ \mu_{t},R_{t}\right] ^{i}-\left[ R_{t},\mu_{t}\right] ^{i}-\frac {1}{%
p!}\left[ R_{t},\mu_{t}^{p}\right] ^{i} & =-\mathrm{d}R_{t}^{i}.
\end{align}
Then, integrating by parts with respect to d and d$_{t}$, we can explicitly
write the variation of $Q_{q-1}\left( \mu\right) $ as an exact form, giving
place to the anomaly proportional to the standard parameter $\varepsilon^{A}$%
\begin{equation}
\delta_{\text{Standard}}Q_{q-1}\left( \mu\right) =\mathrm{d}%
\omega_{q-2}^{1}\left( \varepsilon^{A},\mu\right) ,
\end{equation}
where%
\begin{equation}
\omega_{q-2}^{1}\left( \varepsilon^{A},\mu\right) =\sum_{m,n}\int_{0}^{1}%
\mathrm{d}t\left( 1-t\right) m\left\{ \left( m-1\right) \left\langle
\varepsilon,\mathrm{d}\left( \mu,R_{t}^{m-2};R_{t}^{n}\right) \right\rangle
+npt^{p-1}\left\langle \varepsilon,\mathrm{d}\left( R_{t}^{m-1};\mu
,R_{t}^{n-1}\right) \right\rangle \right\} .  \label{anom2}
\end{equation}
Eqs. (\ref{anom1}) and (\ref{anom2}) determine the total gauge variation of
the CS form in terms of both parameters $\varepsilon^{i}$ and $%
\varepsilon^{A}$ respectively. Note that the definition of $R_{t}$ is
different in both cases due to the different choices of the homotopic gauge
fields.

\section{Concluding remarks}

In this article we have introduced a gauge invariant density for a
particular FDA, analogous to the Chern--Pontryagin topological invariant for
Lie algebras. Such FDA was introduced and studied on Refs. \cite%
{cast1,cast2,cast3,cast4} and its main feature is the presence of a Lie
subalgebra and only one $p$-form extension through the inclussion of a
non-trivial cocycle, representative of a Chevalley--Eilenberg cohomology
class. Explicit expressions for transgression and CS forms whose integral
representation is given by Eq. (\ref{transgression}) and possible gauge
anomalies given by Eqs. (\ref{anom1}) and (\ref{anom2}) were also found. The
procedure to find those expressions is analogous to the one described in
Refs. \cite{zumnb,zumnc} in the calculation of non-trivial chiral anomalies.
This is, by starting from the invariant density, obtain an integral
expression for the corresponding CS form and finally, through its gauge
variation, obtain the anomaly term. The main difference with respect to the
standard case is the presence of extended indices. Since every expression
contains both type of indices $A$ and $i$, being the first in the Lie
subalgebra and the latter in the extended sector of the FDA, it is not
possible to use the standard mathematical identities of group theory. In
particular, the FDAs invariant tensor $g_{A_{1}\cdots A_{m}i_{1}\cdots i_{n}}
$ does not satisfy the invariant tensor conditions of its Lie subalgebra
unless $n=0$. To perform the calculations, it is therefore necessary to use
its definition to obtain the generalized properties to which the invariant
tensor obeys (\ref{zum1}) - (\ref{zum3}). Eqs. (\ref{anom1}) and (\ref{anom2}%
) are the generalization of the non-abelian anomaly (\ref{chiral}),
including non only one form gauge fields but also a $p$-form. Such extension
of the field content in the theory is made in the same way in which the
gauge symmetry is extended from a Lie algebra to an FDA through the
inclusion of a non-trivial cocycle. Since the extension of the anomaly term
has been found in a geometrical framework, its physical meaning must be
separately studied in order to understand its possible consistency with the
breaking of classical symmetries in QFT.

\section*{Acknowledgements}

The autor would like to thank Laura Adrianopoli, Fabrizio Cordonier-Tello,
Nicol\'{a}s Gonz\'{a}lez, Dieter L\"{u}st, Eduardo Rodr\'{\i}guez, Gustavo
Rubio, Patricio Salgado and Mario Trigiante, for enlightening discussions.
This research was partially funded by the bilateral DAAD-CONICYT grant
62160015. The author acknowledges the support from the Max-Planck-Society.

\appendix

\section{Action for the Maxwell-FDA}

Let us consider an extension of the bosonic Maxwell algebra. We introduce a
set of gauge potentials $\mu=\left( \mu^{A},B^{A}\right) $ whose components
are a one-form $\mu^{A}=\left( e^{a},\omega^{ab},k^{ab}\right) $ and a
three-form in the adjoint representation of Maxwell algebra $B^{A}=\left(
b^{a},b^{ab},B^{ab}\right) $. By including a non-trivial cocycle $%
\Omega=\left( 0,0,\Omega^{ab}\right) $ with 
\begin{equation}
\Omega^{ab}=k_{\text{ \ }c}^{a}k_{\text{ \ }d}^{c}e^{d}e^{b}-k_{\text{ \ }%
c}^{b}k_{\text{ \ }d}^{c}e^{d}e^{a}-2k^{ab}k_{cd}e^{c}e^{d},
\end{equation}
it is possible to introduce the following Maurer--Cartan equations, defining
an algebra that we call Maxwell-FDA 
\begin{align}
\mathrm{d}e^{a}+\omega_{\text{ \ }c}^{a}e^{c} & =0,  \label{mfda1a} \\
\mathrm{d}\omega^{ab}+\omega_{\text{ \ }c}^{a}\omega^{cb} & =0,
\label{mfda1b} \\
\mathrm{d}k^{ab}+\omega_{\text{ \ }c}^{a}k^{cb}-\omega_{\text{ \ }%
c}^{b}k^{ca}+\frac{1}{l^{2}}e^{a}e^{b} & =0,  \label{mfda1c} \\
\mathrm{d}b^{a}+\omega_{\text{ \ }c}^{a}b^{c}+b_{\text{ \ }c}^{a}e^{c} & =0,
\label{mfda1d} \\
\mathrm{d}b^{ab}+\omega_{\text{ \ }c}^{a}b^{bc}-\omega_{\text{ \ }%
c}^{b}b^{ba} & =0,  \label{mfda1e} \\
\mathrm{d}B^{ab}+\omega_{\text{ \ }c}^{a}B^{cb}-\omega_{\text{ \ }%
c}^{b}B^{ca}+b_{\text{ \ }c}^{a}k^{cb}-b_{\text{ \ }c}^{b}k^{ca}+\frac{1}{%
l^{2}}\left( e^{a}b^{b}-e^{b}b^{a}\right) +\Omega^{ab} & =0.
\end{align}
By gauging the algebra we consider non-zero curvatures. For convenience we
denote the $2$-form curvature components as $R^{A}=\left(
R^{a},R^{ab},F^{ab}\right) $ and, since $B^{i}$ is in the adjoint
representation, we denote the $4$-form curvature as $R^{i}\rightarrow
H^{A}=\left( h^{a},h^{ab},H^{ab}\right) $. This allows us to propose the
following gauge invariant $6$-form%
\begin{equation}
\chi_{6}\left( \mu,B\right) =g_{ABC}R^{A}R^{B}R^{C}+g_{AB}R^{A}H^{B}.
\end{equation}
In this case, the invariant tensor conditions (\ref{adj1}) - (\ref{adj3})
are given by%
\begin{align}
g_{AD}C_{BC}^{D}R^{A}R^{B}\varepsilon^{C} & =0,  \label{m-inv1} \\
g_{AD}C_{BC}^{A}R^{B}\varepsilon^{C}H^{D}-g_{AD}C_{BC}^{D}R^{A}\varepsilon
^{B}H^{C} & =0,  \label{m-inv2} \\
g_{AB}C_{CDEF}^{B}R^{A}\varepsilon^{C}R^{D}\mu^{E}\mu^{F} & =0.
\label{m-inv3}
\end{align}
As we have seen, in this case, the condition (\ref{m-inv1}) means that $%
g_{AB}$ is also an invariant tensor of the Lie subalgebra. The second
condition becomes equivalent to the first one, while the third condition is
given by%
\begin{equation}
g_{AD}C_{BC}^{D}+g_{BD}C_{AC}^{D}=0.
\end{equation}
Since $g_{AB}$ is an invariant tensor of the Maxwell algebra, we propose the
usual rank-$2$ invariant tensor\footnote{%
We denote with square brackets the indices in the $k^{ab}$ sector of the
algebra. In this case, it doesn't mean antisymmetrization}%
\begin{align}
g_{ab,cd} & =\alpha_{0}\left( \eta_{ac}\eta_{bd}-\eta_{ad}\eta_{bc}\right) ,
\label{g1} \\
g_{ab,\left[ cd\right] } & =\alpha_{1}\left( \eta_{ac}\eta_{bd}-\eta
_{ad}\eta_{bc}\right) ,  \label{g2} \\
g_{ab,c} & =\alpha_{2}\epsilon_{abc},  \label{g3} \\
g_{a,b} & =\alpha_{1}\eta_{ab},  \label{g4}
\end{align}
where $\alpha_{0}$, $\alpha_{1}$ and $\alpha_{2}$ are arbitrary constants.
However, this tensor still has to verify Eq. (\ref{m-inv3}) in order to be
an invariant tensor of the whole FDA. By imposing that condition, we find $%
\alpha_{1}=\alpha_{2}=0$. The rank-$2$ invariant tensor of Maxwell-FDA is
then given by (\ref{g1}).

From (\ref{transgression}) we can write down a Chern--Simons action given by%
\begin{equation}
Q_{5}\left( \mu\right) =3\int_{0}^{1}\mathrm{d}t~g_{ABC}%
\mu^{A}R_{t}^{B}R_{t}^{C}+g_{AB}\int_{0}^{1}\mathrm{d}t~\left(
B^{A_{1}}H_{t}^{B}+R_{t}^{A}B^{B}\right) .  \label{cs-m}
\end{equation}
However, in order to obtain a simpler expression, it is convenient to
introduce a new set of gauge fields and use the triangle relation given by
Eq. (\ref{tri}) in Appendix B. We define $\bar{\mu}=\left( \bar{\mu}^{A},%
\bar {B}^{A}\right) $ whose components are given as follows%
\begin{align}
\bar{\mu}^{A} & =\left( 0,\omega^{ab},k^{ab}\right) , \\
\bar{B}^{A} & =\left( 0,b^{ab},B^{ab}\right) .
\end{align}
Then we can write the CS form (\ref{cs-m}) in terms of a transgression form,
another CS form and total derivatives%
\begin{equation}
Q_{5}\left( \mu\right) =Q_{5}\left( \mu,\bar{\mu}\right) +Q_{5}\left( \bar{%
\mu}\right) +\text{total derivative.}
\end{equation}
From Eq. (\ref{transgression}) we know that the explicit expression for $%
Q_{5}\left( \mu,\bar{\mu}\right) $ is%
\begin{equation}
Q_{5}\left( \mu,\bar{\mu}\right) =\int_{0}^{1}\mathrm{d}t\left(
3g_{ABC}\left( \mu^{A}-\bar{\mu}^{A}\right) R_{t}^{B}R_{t}^{C}+g_{AB}\left[
\left( \mu^{A}-\bar{\mu}^{A}\right) H_{t}^{B}+R_{t}^{A_{1}}\left( B^{B}-\bar{%
B}^{B}\right) \right] \right) .
\end{equation}
In this case, the homotopic gauge fields are given by $\mu_{t}=\bar{\mu }%
+t\left( \mu-\bar{\mu}\right) $ where indexwise $\mu_{t}$ is given by 
\begin{equation}
\mu_{t}^{A}=\left( te^{a},\omega^{ab},k^{ab}\right) ,\text{ \ \ \ \ }\mu
_{t}^{i}=\left( tb^{a},b^{ab},B^{ab}\right) .
\end{equation}

On the other hand, the CS form $Q_{5}\left( \bar{\mu}\right) $ is given by%
\begin{equation}
Q_{5}\left( \bar{\mu}\right) =\int_{0}^{1}\mathrm{d}t~g_{AB}\left( \bar {\mu}%
^{A}\bar{H}_{t}^{B}+\bar{R}_{t}^{A}\bar{B}^{i}\right) ,
\end{equation}
with the homotopic gauge field $\bar{\mu}_{t}=t\bar{\mu}=\left( t\bar{\mu }%
^{A},t\bar{B}^{B}\right) $ and $\bar{R}_{t}=\left( \bar{R}_{t}^{A},\bar {H}%
_{t}^{B}\right) $ being its corresponding curvature. Using the rank-$3$
invariant tensor of the Maxwell algebra $g_{ab,cd,e}=\epsilon_{abcde}$ and
the rank-$2$ invariant tensor (\ref{g1}) for the extended sector, we obtain
the following CS form%
\begin{equation}
Q_{5}\left( \mu\right) =\frac{3}{4}\epsilon_{abcde}R^{ab}R^{cd}e^{e}+\frac{%
\alpha_{0}}{2}\left( \omega^{ab}h_{ab}+R^{ab}b_{ab}\right) .
\end{equation}

The first term corresponds to the usual CS form invariant under the
transformations of Maxwell Lie algebra. The second term extends the
Lagrangian, including the three-form gauge field without breaking the
invariance under transformations of the FDA and modifying the resulting
dynamics in the corresponding theory. However, it is important to note that
the cocycle is not present in the Lagrangian. Since the only non-vanishing
component of the invariant tensor in the extended sector of the algebra is
given by (\ref{g1}), the $B^{ab}$ field is not present in the CS form. To
find an example of a CS Lagrangian invariant under the transformations of an
FDA and involving a non-trivial cocycle in the transformations is still an
open problem.

\section{Subspace separation method}

Extended CS and transgression forms satisfy similar invariance conditions
that their standard versions. There is also a triangle relation between them
that can be explicitly found using the extended Cartan homotopy formula
(ECHF) \cite{mss,subs,tf}.

Let us consider a set of $r+2$ composite gauge connections $\left\{ \mu
_{J}=\left( \mu_{J}^{A},\mu_{J}^{i}\right) \right\} _{J=0}^{r+1}$ defined on
a fiber bundle over $M$ and a $\left( r+1\right) $ dimensional simplex $%
T_{r+1}$ with $r+2$ parameters $t_{J}\in\lbrack0,1]$, satisfying the
constraint $\sum_{J}t_{J}=1$. It is possible to defined an homotopic
connection $\mu_{t}=\sum_{J}t_{J}\mu_{J}$ whose components transform
according (\ref{d1}, \ref{d2}). The ECHF is given by 
\begin{equation}
\int_{\partial T_{r+1}}\frac{\ell_{t}^{s}}{s!}\chi=\int_{T_{r+1}}\frac {%
\ell_{t}^{s+1}}{\left( s+1\right) !}\mathrm{d}\chi+\left( -1\right)
^{s+q^{\prime}}\mathrm{d}\int_{T_{r+1}}\frac{\ell_{t}^{s+1}}{\left(
s+1\right) !}\chi,  \label{cehf}
\end{equation}
where, in this case, $\chi$ represents any polynomial in the forms $\left\{
\mu_{t}^{A},\mu_{t}^{i},R_{t}^{A},R_{t}^{i},\mathrm{d}_{t}\mu_{t}^{A},%
\mathrm{d}_{t}\mu_{t}^{i},\mathrm{d}_{t}R_{t}^{A},\mathrm{d}%
_{t}R_{t}^{i}\right\} $ which is a $q$-form on $M$ and a $q^{\prime}$-form
on $T_{r+1}$, with $q\geq s$ and $s+q^{\prime}=r$. The exterior derivatives
on $M$ and $T_{r+1}$ are denoted respectively by d and d$_{t}$ and the
homotopy operator $\ell_{t}$, which now is defined with respect to the
variations along the simplex, maps mixed differential forms according to%
\begin{gather}
\ell_{t}:\Lambda^{a}\left( M\right) \times\Lambda^{b}\left( T_{r+1}\right)
\rightarrow\Lambda^{a-1}\left( M\right) \times\Lambda^{b+1}\left(
T_{r+1}\right) , \\
\ell_{t}\mu_{t}=0,\text{ \ \ \ \ }\ell_{t}R_{t}=\mathrm{d}_{t}\mu_{t}
\end{gather}%
i.e., $\ell_{t}$ increases the order of the differential form on $T_{r+1}$
and decreases the order on $M$ while it satisfies Leibniz's rule as well as
d and d$_{t}$. Note that the ECHF is different for any value of $s$.
However, its allowed values are $s=0,...q$. As happens in the case studied
in Section 6, the operators d, d$_{t}$ and $\ell_{t}$ define a graded
algebra given by \cite{mss} 
\begin{align}
\mathrm{d}^{2} & =0,\text{ \ \ }\mathrm{d}_{t}^{2}=0,\text{ \ \ }\left\{ 
\mathrm{d},\mathrm{d}_{t}\right\} =0, \\
\left[ \ell_{t},\mathrm{d}\right] & =\mathrm{d}_{t},\text{ \ \ }\left[
\ell_{t},\mathrm{d}_{t}\right] =0.
\end{align}
In this case, we consider the closed polynomial $\chi=\chi_{q}\left( \mu
_{t}\right) $, reducing the ECHF to%
\begin{equation}
\int_{\partial T_{s+1}}\frac{\ell_{t}^{s}}{s!}\chi_{q}=\left( -1\right) ^{s}%
\mathrm{d}\int_{T_{s+1}}\frac{\ell_{t}^{s+1}}{\left( s+1\right) !}\chi_{q}.
\label{cehfc}
\end{equation}
By setting $s=0,$ the homotopic connection is given by $\mu_{t}=\left(
\mu_{t}^{A},\mu_{t}^{i}\right) =\mu_{0}+t\left( \mu_{1}-\mu_{0}\right) $ and
then Eq. (\ref{cehfc}) reproduces the generalized Chern--Weil theorem%
\begin{align}
\chi_{q}\left( \mu_{1}\right) -\chi_{q}\left( \mu_{0}\right) & =\mathrm{d}%
\sum_{m,n\in q\left( p\right) }g_{A_{1}\cdots A_{m}i_{1}\cdots
i_{n}}\int_{T_{1}}\left\{ m\left( \ell_{t}R_{t}^{A_{1}}\right)
R_{t}^{A_{1}}\cdots R_{t}^{A_{m}}R_{t}^{i_{1}}\cdots R_{t}^{i_{n}}\right. 
\notag \\
& \left. +nR_{t}^{A_{1}}\cdots R_{t}^{A_{m}}\left(
\ell_{t}R_{t}^{i_{1}}\right) R_{t}^{i_{2}}\cdots R_{t}^{i_{n}}\right\} .
\label{p0}
\end{align}
For $s=1$ we have a three-dimensional simplex, the homotopic gauge field is
given by $\mu_{t}=t^{0}\left( \mu_{0}-\mu_{1}\right) +t^{2}\left( \mu
_{2}-\mu_{1}\right) +\mu_{1}$ and Eq. (\ref{cehfc}) takes the form 
\begin{equation}
\int_{\partial T_{2}}\ell_{t}\chi_{q}\left( \mu_{t}\right) =-\mathrm{d}%
\int_{T_{2}}\frac{\ell_{t}^{2}}{2}\chi_{q}\left( \mu_{t}\right) .  \label{p2}
\end{equation}
Performing the integration directly over $T_{2}$ and $\partial T_{2}$ we
find a triangle relation for the generalized transgression forms%
\begin{equation}
Q_{q-1}\left( \mu_{1},\mu_{2}\right) -Q_{q-1}\left( \mu_{0},\mu_{2}\right)
+Q_{q-1}\left( \mu_{0},\mu_{1}\right) =\mathrm{d}Q_{q-2}\left(
\mu_{2},\mu_{1},\mu_{0}\right) .  \label{tri}
\end{equation}
Note that if we choose $\mu_{0}=\left( \mu_{0}^{A},\mu_{0}^{i}\right)
=\left( 0,0\right) $ we obtain an expression that relates the transgression
form with two CS forms and a total derivative. This equation is useful in
explicit calculations of CS and transgression actions.

\end{document}